\documentclass[prl, aps, twocolumn, superscriptaddress, nofootinbib, tightenlines, nobibnotes, showpacs, 10pt]{revtex4-1}
\usepackage{natbib}
\usepackage{slashed}
\usepackage{graphicx}
\usepackage{subfigure}
\usepackage[usenames, dvipsnames]{color}
\usepackage{graphics}
\usepackage{hyperref}
\usepackage{bm}
\usepackage{amsmath}
\usepackage{color}
\usepackage{amsfonts}
\hypersetup{backref,
colorlinks=true,
linkcolor=blue,
linktoc=page,
citecolor=blue,
urlcolor=blue}

\everymath{\displaystyle}

\begin{document}

\title{Axion production in unstable magnetized plasmas}

\author{J. T. Mendon\c ca}
\affiliation{Instituto de Plasmas e Fus\~ao Nuclear, Lisboa, Portugal}
\affiliation{Instituto Superior T\'ecnico, Lisboa, Portugal}

\author{J. D. Rodrigues}
\affiliation{Instituto de Plasmas e Fus\~ao Nuclear, Lisboa, Portugal}
\affiliation{Instituto Superior T\'ecnico, Lisboa, Portugal}
\affiliation{Physics Department, Blackett Laboratory, Imperial College London, United Kingdom}

\author{H. Ter\c cas}
\email{hugo.tercas@tecnico.ulisboa.pt}
\affiliation{Instituto de Plasmas e Fus\~ao Nuclear, Lisboa, Portugal}
\affiliation{Instituto Superior T\'ecnico, Lisboa, Portugal}

\begin{abstract}

Axions, the hypothetical particles restoring the charge-parity symmetry in the strong sector of the Standard Model, and one of the most prone candidates for dark matter, are well-known to interact with plasmas. In a recent publication [Phys. Rev. Lett. {\bf 120}, 181803 (2018)], we have shown that if the plasma dynamically responds to the presence of axions, then a new quasi-particle (the axion plasmon-polariton) can be formed, being at the basis of a new generation of plasma-based detection techniques. In this work, we exploit the axion-plasmon hybridization to actively produce axions in streaming magnetized plasmas. The produced axions can then be detected by reconversion into photons, in a scheme that is similar to the light-shining-a-wall experiments. 
\end{abstract}
\maketitle


{\it Introduction.} Axions and axion-like particles are hypothetical particles (ALPs) that have been proposed to solve the strong CP problem \citep{pendlebury_2015, jaeckel_2010, ringwald_2012}. At the origin of the latter, is the fact that non-perturbative (instanton) effects force the QCD
Lagrangian to contain a total derivative with an arbitrary parameter (an angle $\theta$) which does not vanish at infinity, therefore violating the CP symmetry. This is in blatant contradiction with the fact that strong interactions conserve CP \citep{kim_2010}. Strong bounds on the neutron electric dipole moment imply that $\theta\lesssim 10^{-9}$ for the QCD to be compatible with the experiments \citep{pendlebury_2015}. A first, dynamical mechanism allowing $\theta\rightarrow0$ was put forward by Peccei and Quinn \citep{quinn_1977}, with the axion being later identified as the Goldstone boson associated to the spontaneous symmetry breaking of the continuous Peccei-Quinn $U(1)_{\rm PQ}$ symmetry \citep{weinberg_1978, wilczek_1978}. 

Axions and ALPs are predicted to have an extremely small mass (possibly in the meV range) and couple very weakly to ordinary matter. For that reason, ALPs became appealing candidates (arguably, the most well theoretically motivated) to fix the dark matter puzzle as well \cite{PhysRevLett.51.1415, PhysRevLett.104.041301}. Many facilities have been built with the goal of observing axion or ALP signatures, both based on laboratory and astrophysical observations \cite{PhysRevLett.94.121301, PhysRevLett.98.201801,PhysRevLett.118.261301}. However, given the smallness of the axion-photon coupling, testing the axion is  difficult, rendering most of the experimental observations inconclusive. Telescope experiments, such as CAST - the most recent results establishing $g<0.66\times 10^{-10}$ GeV$^{- 1}$ for $m_\varphi < 0.02$ eV at the $2\sigma$ level \cite{Collaboration2017} -, and ADMX \cite{asztalos_2001, asztalos_2010, admx_2018}, IAXO \cite{vogel_2015} and MADMAX \cite{caldwell_2017}, investigating more precise regions of the QCD axion parameter space, are designed to probe axions produced by astrophysical objects. By construction, these experiments rely on a {\it passive} approach, in the sense that no axion production is envisaged. It is therefore desirable to look for alternatives, where axions could be {\it actively} produced in the lab. This motivation is at the basis of the ``light shining throw a wall" (LSW) strategy \citep{friederike_2014}, such as those implemented by ALPS II \cite{baehre_2013} and OSQAR \citep{osqar_2008}, using near infrared and visible light, STAX \citep{capparelli_2016} and CROWS \cite{betz_2013}, using sub-THz and microwave radiation.\par     
\begin{figure}[t!]
\includegraphics[scale=0.85]{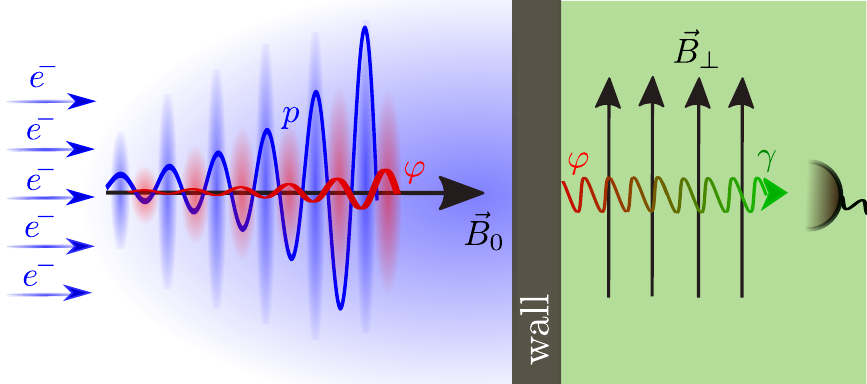}
\caption{(color online). Sketch of a ``plasma shining a wall" (PSW) experiment based on the beam-plasma instability. A collimated electron beam (blue arrows) is injected into a cold plasma (blue shadow), leading to the growth of plasmons ($p$). The longitudinal magnetic field $\mathbf {B}_0$ then converts the plasmons into axions or ALPs ($\varphi$). The plasma and its radiation are blocked by a wall. The axions passing the wall are finally converted back into photons ($\gamma$) in the transverse magnetic field $\mathbf {B}_\perp$, being probed by a single-photon microwave detectors.}
\label{fig_scheme}
\end{figure}
One important limitation of the previous LSW schemes is the extremely low value of the axion-photon (and vice-versa) conversion probabilities, a fact than can be somehow circumvented by allowing axion conversion to take place in a plasma \citep{raffelt_1988}. Actually, there is a recent hype around plasmas in the context of particle physics. The wakefield acceleration paradigm, for example, has gained much breath as it reveals to be an efficient way to accelerate particles \citep{ee183cdbe0634a8dbaabf276f03a1d91, Geddes2004, Faure2004}, as recently demonstrated by the latest experiments by the AWAKE collaboration \cite{adli_2018}. Interestingly, recent theoretical studies have pointed out that such wakefields could ultimately be used to produce ALPs in the lab \citep{1710.01906, 1751-8121-43-7-075502,1751-8121-49-38-385501}, and that petaWatt lasers could also do the job \cite{0295-5075-79-2-21001}. Plasmas are also playing a prominent role in axion astrophysics, as they have been put forward as veicules for efficient axion-photon conversion in the atmosphere of magnetars \citep{pshirkov_2009, hook_2018,srimoyee_2018}.     

In this Letter, we show that axions and axion-like particles can be actively produced in an unstable magnetized plasma, putting forward the physical principle for a ``plasma shining a wall" (PSW) strategy. If an energetic electron beam is injected in the plasma, unstable electron waves, or plasmons, are produced. This effect is dubbed in the literature as the beam-plasma instability \cite{9781475704617, asseo_1980}. The growing plasmon perturbation then provides the energy to the growth of axions. A schematic representation of the process is depicted in Fig. \ref{fig_scheme}. In the absence of axions, the plasmons are insensitive to the magnetic field; however, if axions exist, they admix with the plasmons, leading to the formation of a hybrid quasi-particle, the axion-plasmon polariton \citep{tercas_2018}. As such, if the plasmons become dynamically unstable, their small axion component will also grow, leading to an efficient axion production in laboratory conditions. As a matter of fact, plasmon-axion mixing (differing from photon-axion mixing in plasmas) has been first considered in Ref. \cite{das_2008}, although no physical consequences have been exploited there. Our estimates based on realistic experimental conditions show that a remarkably high plasmon-axion conversion probability can be achieved, as a consequence of the beam-plasma instability. We predict a detectable photon signal for the axions passing the wall. Some implications in the radio signals emitted by pulsars are also discussed.

\begin{figure}[t!]
\flushleft
\includegraphics[scale=0.5]{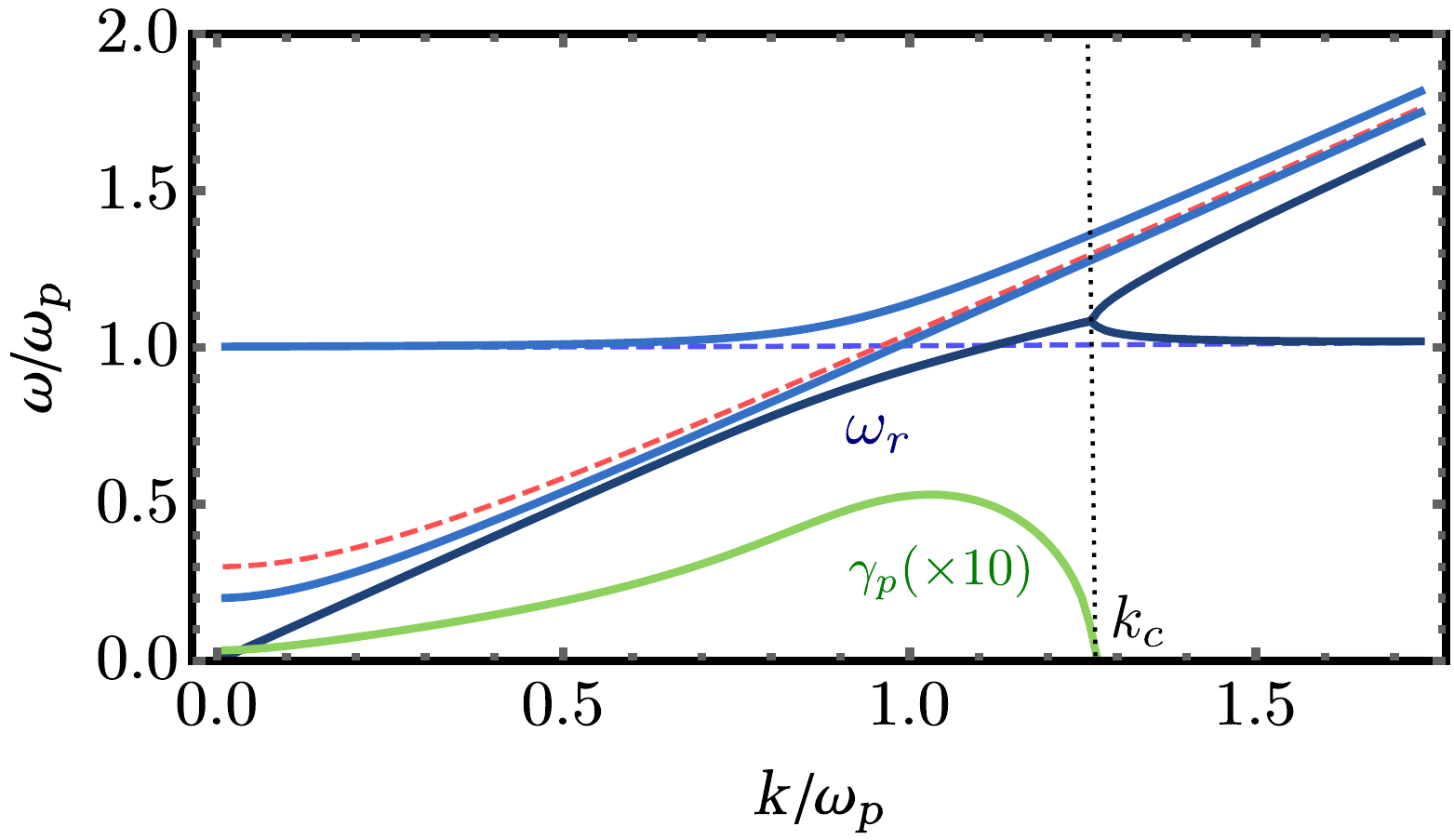}
\includegraphics[scale=0.5]{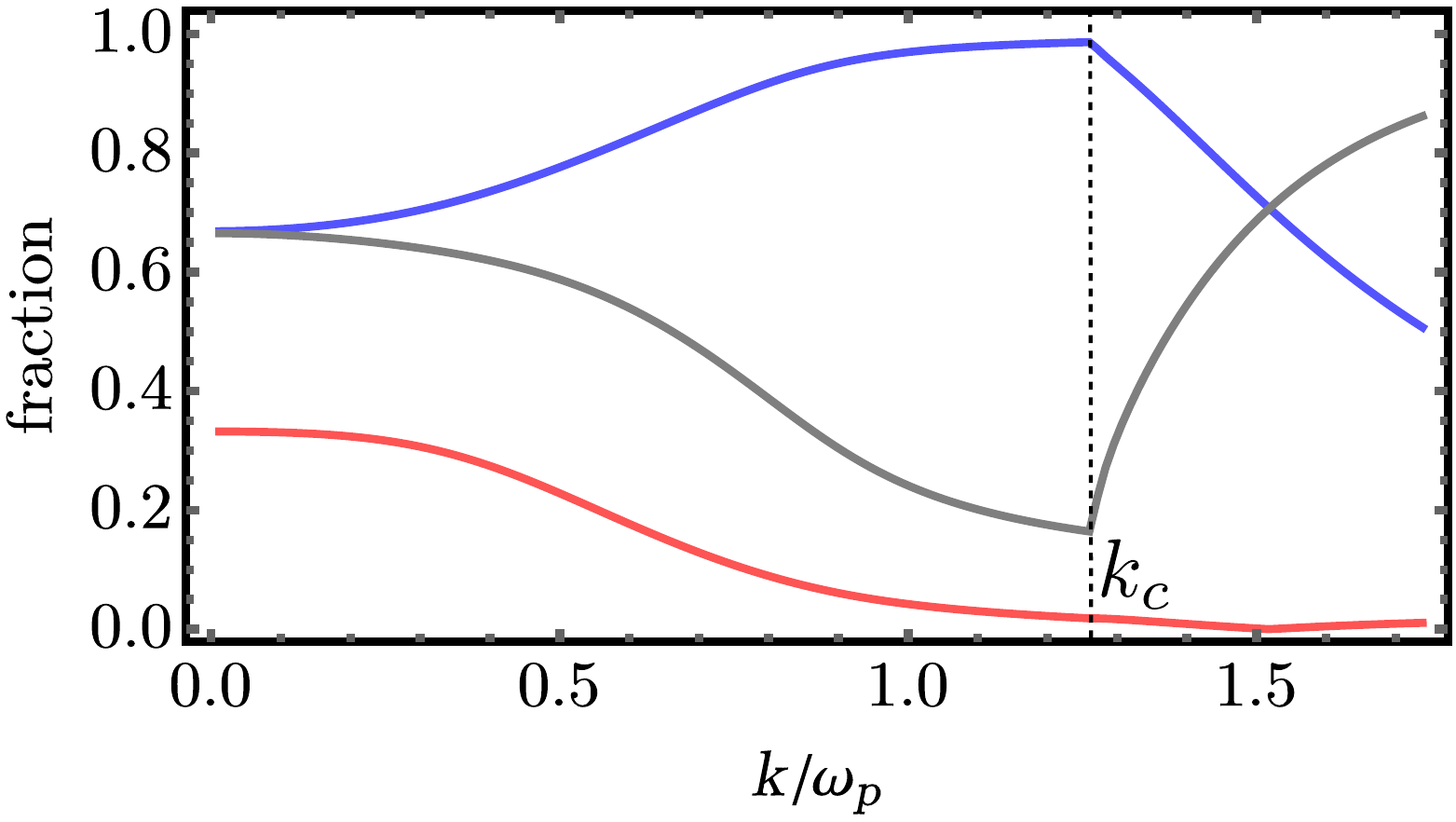}
\caption{(color online). Top panel: Dispersion relation of the axion-plasmon polariton in the streaming instability situation. The dashed lines are the bare dispersions. Axion (red), plasmon (blue) and electron beam (grey). The black solid lines depict the real part of the modes, while the green line represents the imaginary part of the coalesced plasma-beam mode. For illustration, we have set $f=0.1$, $m_\varphi=0.3 \omega_p$, $u_0=0.99$ and the exaggerated value $\Omega=0.1 \omega_p$. Bottom panel: the axion (red curve), plasma (blue curve) and beam (grey curve) fractions in the unstable mode, as obtained by extracting the eigenvalues of Eq. \eqref{eq_eigenvalue}.}
\label{fig_dispersion}
\end{figure}

{\it Beam-plasma instability in magnetized plasmas.} The minimal electromagnetic theory accommodating the axion-photon coupling can be constructed as follows ($\hbar=c=1$) \citep{ visinelli_2013, wilczek_1987} 
\begin{equation}
\mathcal{L}=-\frac{1}{4}F_{\mu\nu}F^{\mu\nu}-A_\mu J^\mu_e+\mathcal{L}_\varphi +\mathcal{L}_{\rm int}, \label{eq_lagrange}
\end{equation}
where $F_{\mu \nu}=\partial_\mu A_\nu-\partial_\nu A_\mu$ is the electromagnetic (EM) tensor, $J_e^{\mu}$ is the electron four-current, and $\mathcal{L}_\varphi=\partial_\mu \varphi^*\partial^\mu \varphi-m_\varphi^2 \vert \varphi \vert^2$ is the axion Lagrangian (with $\varphi$ denoting the axion field). For the QCD axion, $m_\varphi=\sqrt{z}f_\pi m_\pi/f_\varphi$, where $z=m_u/m_d$ is the up/down mass ratio, and $f_{\varphi(\pi)}$ is the axion (pion) decay constant \cite{quinn_1977, weinberg_1978}. Upon integration of the anomalous of the axion-gluon triangle, one obtains $\mathcal{L}_{\rm int}=-(g/4)F_{\mu \nu}\tilde F^{\mu \nu}$, where $\tilde F^{\mu \nu}=\epsilon^{\mu \nu \alpha \beta} F_{\alpha \beta}$ denotes the dual EM tensor and $g$ is the axion-photon coupling. Although motivated for the QCD axion, the remainder of the paper is valid for any ALP. From Euler-Lagrange equations, one obtains the Maxwell's equations \citep{tercas_2018}, in particular the Poisson equation
\begin{eqnarray}
\begin{array}{c}
\bm \nabla\cdot\left({\bf E}+g\varphi {\bf B}\right)=\rho, 
\end{array}
\label{eq_maxwell}
\end{eqnarray}
and the Klein-Gordon equation describing the axion field
\begin{equation}
\left(\square +m_\varphi^2\right)\varphi= g {\bf E}\cdot {\bf B},
\label{eq_KG}
\end{equation}
with $\square= \partial_t^2-\nabla^2$ denoting the d'Alembert operator. In the situation of an electron beam propagating inside the plasma, $\rho=e(n_i-n_e-n_b)$, where $n_i$, $n_e$ and $n_b$ respectively represent the ion, electron and beam densities. As we are interested in electron plasma waves only, we can assume the ions to be immobile. Thus, the equations governing the dynamics of the plasma and beam electrons are given by
\begin{equation}
\frac{\partial n_\alpha}{\partial t}+\bm \nabla\cdot \left(n_\alpha \mathbf{u}_\alpha \right)=0,
\label{eq_continuity}
\end{equation}
with $\alpha=\{e,b\}$, and
\begin{eqnarray}
\left(\frac{\partial}{\partial t}+\mathbf{u}_\alpha\cdot \bm \nabla\right)\mathbf{u}_\alpha=-\frac{e}{\gamma_\alpha m_e}\left(\mathbf{E}+{\bf u}\times {\bf B}\right),
\label{eq_force}
\end{eqnarray}
where $\gamma_\alpha=(1-u_\alpha^2)^{-1/2}$ is the Lorentz factor. In the following, we will consider the plasma electrons to be initially at rest ($\gamma_e\simeq 1$), while the beam electrons propagate with velocity ${\bf u}_0$. We are interested in describing the electrostatic perturbations along a static, homogeneous magnetic field ${\bf B}=B_0 {\bf e}_z$. As such, owing to the quasi-neutrality condition of the plasma, we perturb the densities as $n_e= n_0 + \tilde n_e$ and $n_b= f n_0+\tilde n_b$ (here, $f$ is the fraction of the electrons in the beam), and the axion field as $\varphi=\tilde \varphi$ (neglecting the presence of a v.e.v., $\varphi_0=0$) to obtain
\begin{equation}
\begin{array}{c}
\frac{\partial ^2}{\partial t^2}\tilde n_e - \frac{e n_0}{m_e}   \frac{\partial E}{\partial z}=0, \\
\left(\frac{\partial}{\partial t} +  u_0 \frac{\partial}{\partial z} \right)^2 \tilde n_b +  \frac{f}{\gamma_0}\frac{e n_0}{m_e}  \frac{\partial E}{\partial z}=0,\\
\left(\frac{\partial^2}{\partial t^2}-\nabla^2 + M_\varphi^2 \right)\frac{\partial \tilde \varphi}{\partial z}-g B_0 \frac{\partial E}{\partial z}=0.
\end{array}
\label{eq_eigenvalue}
\end{equation}
After Fourier transforming, this allows us to write Eq. (\ref{eq_maxwell}) as $i k[\epsilon(k, \omega)E]=0 $, where
\begin{equation}
\begin{array}{l r}
\epsilon(k, \omega)=  1-\frac{\omega_p^2}{\omega^2}-\frac{f}{\gamma_0}\frac{\omega_p^2}{(\omega-k u_0)^2}& \\
-\frac{\Omega^4}{\omega^2(\omega^2-\omega_\varphi^2)}-\frac{f}{\gamma_0}\frac{\Omega^4}{(\omega-k u_0)^2(\omega^2-\omega_\varphi^2)}&
\end{array}
\label{eq_dispersion}
\end{equation}
is the dielectric permittivity, $\omega_p=\sqrt{e^2n_0/(m_e)}$ is the plasma frequency, and $\omega_\varphi^2=M_\varphi^2+k^2$, with $M_\varphi=\sqrt{m_\varphi^2+g^2B_0^2}$ being the axion effective mass in the plasma. Here,
\begin{equation}
\Omega=\sqrt{gB_0 \omega_p}\sim 2\pi\times (1.2 {~\rm Hz})\sqrt{\frac{g\times 10^{13}}{ {~\rm GeV}^{-1}}\frac{B_0}{ {\rm T}}\frac{\omega_p}{\rm  GHz}}
\end{equation}
is the axion-plasmon coupling parameter (Rabi frequency). In the absence of the beam ($f=0$), Eq. \eqref{eq_dispersion} yields the lower (L) and (U) polariton modes \cite{tercas_2018}
\begin{equation}
\omega_{\rm L(U)}^2=\frac{1}{2}\left( \omega_\varphi^2 +\omega_p^2 \mp \sqrt{(\omega_\varphi^2-\omega_p^2)^2+4\Omega^4}\right).
\label{eq_polariton}
\end{equation}  
Conversely, in the absence of axions, Eq. \eqref{eq_dispersion} describes the celebrated beam-plasma instability \cite{anderson_2001, 9781475704617}. For modes satisfying the condition $k\leq k_c$, with
\begin{equation}
k_c =\frac{\omega_p}{u_0}\left(1+\nu^{1/3}\right)^{3/2}, \quad (\nu=f/\gamma_0)
\end{equation}
being the cut-off wavevector, the plasma and the beam (with dispersion $\omega=\sqrt{\nu}\omega_p +u_0 k$) modes coalesce and the resulting dispersion relation becomes complex. In the unstable region, the dispersion relation of the plasma reads $\omega\simeq \omega_r+i\gamma_p$, where $\omega_r=u_0 k (1-\nu^{2/3})$ and $\gamma_p$ is the instability growth rate \cite{9781475704617, anderson_2001}
\begin{equation}
\gamma_p=\frac{\nu^{2/3}}{\sqrt{3}(1+\nu^{4/3})^{5/2}}\frac{u_0^2 k^2}{\omega_p}\left[\frac{\omega_p^2}{u_0^2k^2}\left(1+\nu^{4/3}\right)^3-1\right]^{1/2}.
\label{eq_gamma_p}
\end{equation}
The most unstable mode, occurring at $k\simeq \omega_p/u_0 $, grows at the rate $\gamma_p^{\rm max}\simeq 0.69 \nu^{2/3}\omega_p$. These features are depicted in Fig. \ref{fig_dispersion} a). \par
{\it Plasmon-axion conversion.} Given the smallness of $\Omega$, the instability does not change noticeably in the perspective of the plasma, and therefore the discussion above holds even in the presence of axions. However, and more crucially, the small fraction of axion that participates in the beam-plasma dynamics leads to the production of axions. The fractions (i.e. the eigenvectors) can be determined by solving the eigenvalue problem in Eq. \eqref{eq_eigenvalue} numerically, as illustrated in Fig. \ref{fig_dispersion} b). The axion production mechanism can thus be understood as follows: the beam transfers energy to the plasma, which becomes unstable; then, the magnetic field mixes the axion and the plasmon modes, allowing the latter to be converted into the former. To estimate this, we notice that the coupling between the axion and the plasma is much larger than that with the beam, resulting in the separation of scales $\omega_p^2\gg \Omega^2 \gg \nu \Omega^2$. Under this conditions, we can solve Eqs. \eqref{eq_eigenvalue} numerically to compute the plasmon-axion conversion probability in the beam-plasma configuration. In the quasi-linear diffusion regime, allowing to accommodate the instability saturation by substituting $\gamma_p\rightarrow \gamma_p[1- 9\omega_p^4 n(t)^2/(8\gamma_p^4 n_0^2)]$ in the eigenvalue problem \cite{oneil_1971,sharma_1976}, we obtain the following piecewise function (see \citep{supp} for details)
\begin{equation}
P_{p\rightarrow \varphi}=\left\lbrace
\begin{array}{c}
e^{2\gamma_p t}P_{p\rightarrow \varphi}^{\rm osc}, 	\quad t\leq \tau_{\rm sat} \\\\
e^{2\gamma_p \tau_{\rm sat}}P_{p\rightarrow \varphi}^{\rm osc}, \quad t>\tau_{\rm sat} 
\end{array} 
\right. 
\label{eq_prob1}
\end{equation}
where $\tau_{\rm flight}=L/v_\varphi$ (with $v_\varphi=\partial_k\omega_\varphi$) is the axion time-of-flight in a plasma column of size $L$, and $\tau_{\rm sat}\sim \nu^{1/3}\omega_p/\gamma_p^2$ is the instability saturation time. Here,
\begin{equation}
P_{p\rightarrow \varphi}^{\rm osc}=\frac{g^2B_0^2 \sin^2 \left[\frac{t}{2}\sqrt{g^2B_0^2-(\omega_r-\omega_\varphi)^2}\right]}{4\left[g^2B_0^2-(\omega_r-\omega_\varphi)^2-\gamma_p^2\right]}
\end{equation}
is the oscillating probability in the plasma \cite{supp}. For a discharge plasma column of size of $L\sim 3.15$ m and plasma frequency $\omega_p\sim 2\pi \times 1$ GHz in a magnetic field of $B_0 \sim 1$ T, with a growth rate of $\gamma_p\sim 10^{-2}\omega_p$ at resonance, and taking $g\sim 10^{-14}$ GeV$^{-1}$, we obtain $P_{p\rightarrow \varphi}\sim 10^{-21}$ for the most unstable mode, $k\sim \omega_p/u_0$. This happens for sufficiently light axions $m_\varphi\lesssim 0.1\omega_p$, for which $\tau_{\rm flight}<\tau_{\rm sat}$. For higher values of the mass, $\tau_{\rm flight}>\tau_{\rm sat}$, and the saturation probability can go up to $10^{-16}$ \cite{supp}. \par

\begin{figure}[t!]
\flushleft
\includegraphics[scale=0.68]{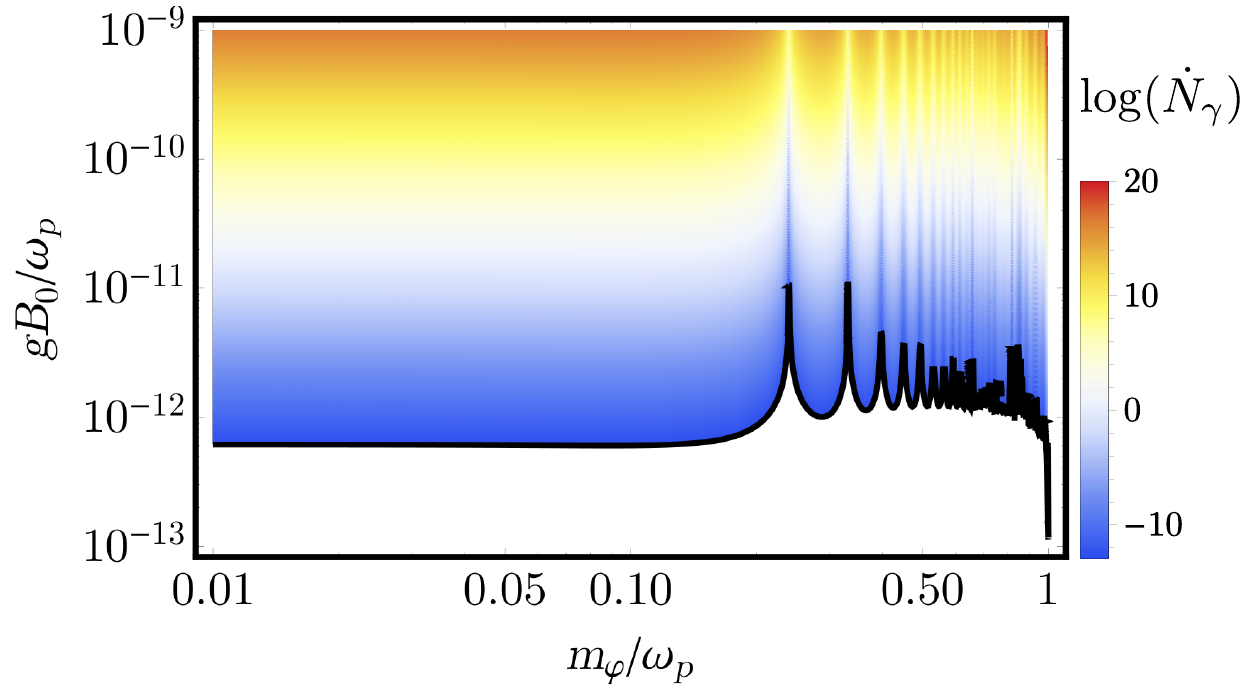}
\caption{(color online). Photon counts per second in the PSW scheme, computed for a regeneration path of $d=10$ m in a transverse magnetic field $B_\perp=$10 T. The black line depicts the estimated sensitivity after an observation time of $100$ hours. Here, we have considered a 1 m$^3$ plasma in a cylindrical column of $L=3.15$ m with plasma frequency $\omega_p=2\pi\times 1 $ GHz in $B_0=1$ T magnetic field. The electron beam concentration is $f=0.1$, with an energy of $E_b=3$ MeV and a repetition rate $R_b=100$ Hz.}
\label{fig_rates}
\end{figure}

The axions resulting from the PSW experiment above can then be sent into a regeneration chamber and be eventually converted into photons, similarly to what is done in the LSW schemes \citep{friederike_2014}. For that task, we consider a homogeneous, transverse  magnetic field $\mathbf{B}_\perp$ in a cavity of length $d$, for which the corresponding axion-photon conversion rate is given by $P_{\varphi\rightarrow \gamma}\simeq \sin^2\Theta \sin^2(\Delta k d)$, where $\tan(\Theta)=g B_0 \omega/(m_\varphi^2-m_\gamma^2)$ is the mixing angle and $\Delta k=\vert \sqrt{\omega^2-m_\varphi^2}-\sqrt{\omega^2-m_\gamma^2} \vert$ is the axion-photon momentum difference and $m_\gamma$ is the photon mass in the buffer gas \citep{raffelt_1988}. To estimate the order of magnitude of the photon flux at the detector, we relate the energy $\mathcal{E}$ delivered in the plasma by an electron beam of energy $E_b$ and density $f n_0$ to the number of plasmons created, i.e. $\mathcal{E}/V=f n_0 E_b=N_p\omega_p/V=N_{\varphi}\omega_p/(V P_{p\rightarrow \varphi})$, with $N_{p(\varphi)}$ denoting the average number of plasmons (axions). Assuming that the beams can be injected in a plasma at the repetition rate $R_b$, we finally obtain the number of photons detected per unit volume   
\begin{equation}
\frac{\dot N_\gamma}{V}\simeq \frac{R_b}{\omega_p} f n_0 E_b  P_{p\rightarrow \varphi} P_{\varphi \rightarrow \gamma}\epsilon_{\rm det},
\label{eq_counts}
\end{equation}
where $\epsilon_{\rm det}$ is the efficiency of the single-photon detector. For the conditions discussed above, and assuming a magnetic field of $B_\perp = 10$ T to be homogeneous in a cavity of length $d=10$ m \cite{friederike_2014}, a detector of efficiency $\epsilon_{\rm}=0.9$ is able to detect resonant axions $\omega_\varphi \sim \omega_p$ after an observation time of $100$ hours. This remarkable property of axion amplification in plasmas make PSW experiments good candidates to probe the QCD axion. A theoretically estimated sensitivity in the $g-m_\varphi$ plane (normalized to the plasma parameters, for generality) relative the proposed PSW scheme is depicted in Fig. \ref{fig_rates}. In our estimates, we have considered the case $m_\gamma=0$, but the sensitivity can by the improved by introducing a buffer gas in the recombination chamber.  

{\it Axion production in pulsars.} Our findings can also be interesting to identify axion production via plasma instabilities taking place in magnetar magnetic pole caps. As it is known, alongside with the gamma-ray emission taking place in the region of high-density, boosted plasma \citep{comment}, the beam-plasma instability leads to the formation of plasma bunches that generate radio emission via the curvature effect \cite{ochelkov_1984, ruderman_1975}. During this process, the produced axions may be resonantly converted into photons at the radius $r_c$ related to the Goldreich-Julian magnetosphere density \cite{goldreich_1969, melrose_2016}   
\begin{equation}
n_c=\frac{2\pi B_0}{eP}\frac{1}{1-4\pi^2 r_c^2 \sin^2\theta/P},
\end{equation}
where $P$ is the pulsar period and $\theta $ is the polar angle with respect to the rotation axis. For $\theta=90^\circ$, the corresponding plasma frequency is $\omega_p/2\pi \simeq(1.5 \times 10^2~{\rm GHz})\sqrt{(B/10^{14}{~\rm G})(1 {~\rm sec}/P)}$, yielding $\omega_p\sim 2\pi \times 98$ GHz for the SGR J145-2900 magnetar ($P\simeq 3.76$ s, $B_0 \simeq 1.6 \times 10^{14}$ G \cite{kennea_2013, eatough_2013}), a value not too far from the discharge plasma discussed above. 
%
%
%
%
There are two electromagnetic modes propagating in a transverse magnetic field: the ordinary (the O-) mode, with parallel polarization $ {\bf E}\parallel {\bf B}_0$, and the extraordinary (the X-) mode, of perpendicular polarization ${\bf E}\perp {\bf B}_0$ \cite{9781475704617}. From Eq. \eqref{eq_KG}, it is clear that only the former can result from the axion-photon decay process, and satisfies the dispersion relation $\omega^2=\omega_p^2+k^2$.  Since only photons with frequency larger than the $\omega_p$ escape the plasma, and given that the plasma instability terminates at the cut-off frequency $\omega_{c}=\sqrt{m_\varphi^2+k_c^2}$, axion-photon conversion will occur in the range $\omega_p\leq\omega\leq\omega_c$. For resonant conversion, $\omega_p\simeq m_\varphi$, the cut-off frequency reads 
\begin{equation}
\omega_c= \sqrt{m_\varphi^2+k_c^2}\simeq \sqrt{2} \omega_p\left[1+\frac{3}{4}\left(\frac{fm_e}{E_b}\right)^{1/3}\right],
\label{eq_range}
\end{equation}   
valid for relativistic electron beams, $E_b\gg m_e$. Assuming the electron beam to be much more energetic than the electron-positron plasma (making the cold plasma model valid), and taking a beam relativistic factor of $\gamma_0 \sim 10^6$, we estimate a cut-off frequency of $\omega_c\simeq 2\pi\times 137 $ GHz. As such, a signal in the range $98 {~\rm GHz} \leq  \omega/ 2\pi \leq 137$ GHz might be expected for the conditions of the experiment proposed in Refs. \cite{huang_2018, hook_2018}, based on axion dark matter conversion (notice that in our case we do not need a dark matter background). At this stage, however, we can not commit whether or not the axions produced via streaming instability can be detected within the sensitivity of telescopes such as CAST or the Arecibo Telescope for the typical observation periods (this would involve a more detailed calculation of the beam injection rates, intensity, energies etc.), but we anticipate that the narrow spectrum in Eq. \eqref{eq_range} would be a clear signature of this process.

{\it Conclusion.} We have shown that a magnetized plasma can be an active source of axions and axion-like particles. For that, we exploit the beam-stream instability triggered by a monoenergetic electron beam to convert plasmons into axions. The production mechanism is based on the transfer of energy from the electrons to the small axion-plasmon admixture, the later being a consequence of the axion-plasmon polariton coupling occurring in magnetized plasmas \citep{tercas_2018}. An estimation of the sequent conversion of the axion into a photon in a transverse magnetic field suggests that our schemes can compete with some of the existing light-shining-through-a-wall experiments such as ALPS and OSQAR \cite{redondo_2011, alps_2010, osqar_2008}. Based on realistic parameters, we estimate a discharge plasma column of size $L\sim 3$ m, in a magnetic field of the order of 1 T, excited by a 3 MeV-electron beam may be able to produce axions close to the QCD region. Our findings will certainly urge the design of a ``plasma-shining-a-wall'' setup in the near future. On the other hand, given the abundance of astrophysical bodies displaying beam-plasma and beam-beam instabilities, we anticipate that a plethora of new exciting phenomena involving the dynamics of axions in plasma may arise in the near future, adding a new twist to the growing interest on axions and axion-like particles in astrophysics \cite{hook_2018,srimoyee_2018, pshirkov_2009, angelis_2007,tavecchio_2012, galanti_2018}.

\par
The authors acknowledge FCT - Funda\c{c}\~{a}o da Ci\^{e}ncia e Tecnologia (Portugal) through the grant number IF/00433/2015. 

\begin{center}
\textbf{Supplemental Material: Calculation of the conversion probability in the unstable plasma}
\end{center}
\setcounter{equation}{0}
\setcounter{figure}{0}
\setcounter{table}{0}
\setcounter{page}{1}
\makeatletter
\renewcommand{\theequation}{S\arabic{equation}}
\renewcommand{\thefigure}{S\arabic{figure}}

\begin{figure}[t!]
\includegraphics[scale=0.63]{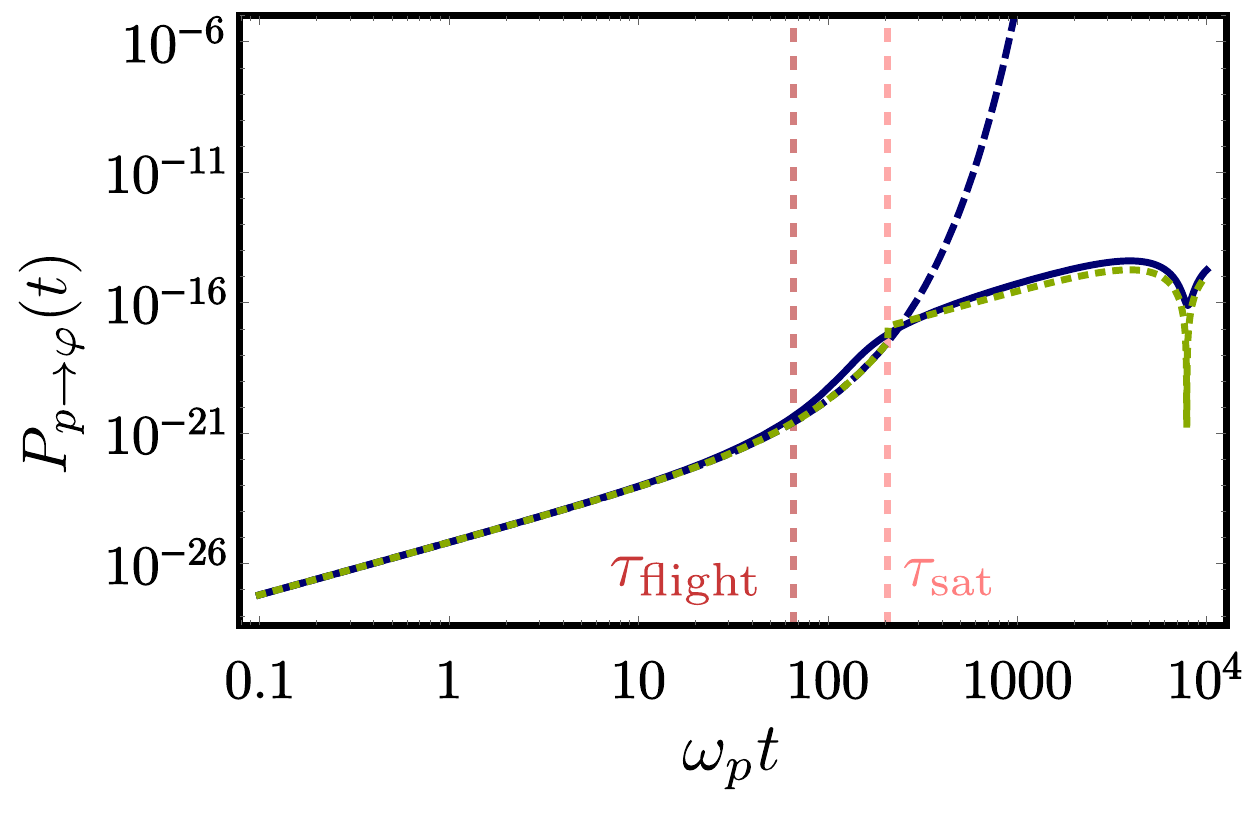}
\caption{Illustration of the plasmon-axion conversion probability for the near-resonant case $\omega_\varphi=0.99\omega_p$. The dashed line corresponds to the linear approximation given in Eq. \eqref{eq_prob_unstable}, while the solid line is the full numerical result (including the plasma saturations). The dotted line corresponds to the piecewise definition in Eq. \eqref{eq_prob_final}, interpolating both the linear and the nonlinear stages of the instability. As one can observe, for light-enough axions ($m_\varphi=0.1\omega_p$), the linear approximation provides an excellent approximation. As previously, we have chosen $\omega_p=2\pi \times 1$ GHz, $g=10^{-13}$ GeV$^{-1}$, $B_0=1$ T,  $f=0.1$ and $u_0=0.99$. The plasma column is set to be $L=3.15$ m, just as in the main text.}
\label{fig_prob_k}
\end{figure}

We start from the dispersion relation appearing in Eq. (9) of the manuscript. Using the plane-wave decomposition
\begin{equation}
\begin{array}{ccc}
n(z,t)&=&\sum_{k} n_k e^{i(\omega t-k z)},\\ 
\quad \varphi(z,t)&=&\sum_{k}  \varphi_k e^{i(\omega t-kz)},
\end{array}
\end{equation}
it can be written in the form
\begin{equation}
\begin{array}{c}
\left(\omega^2-\omega_p^2\right) n_k+ i g B_0 \frac{en_0 k}{m_e}   \varphi_k = 0\\\\
\left(\omega^2-\omega_\varphi^2\right)\varphi_k -i g e B_0 n_k=0,
\end{array}
\label{eq_modes1}
\end{equation} 
where $\omega_\varphi=\sqrt{m_\varphi^2+g^2 B_0^2+k^2}$. We now make use of the secular approximation (also known as rotating-wave approximation) to take the slow varying amplitudes only, by making $$\omega^2-\omega_p^2=(\omega+\omega_p)(\omega-\omega_p)\simeq 2\omega_p(\omega-\omega_p),$$ and similarly for the term $(\omega^2-\omega_\varphi^2)$. Inserting in Eq. (\ref{eq_modes1}), we obtain
\begin{equation}
\begin{array}{c}
(\omega-\omega_p)n_k + i g B_0 \frac{en_0 k}{2\omega_p m_e}\varphi_k=0,\\\\
(\omega-\omega_\varphi)\varphi_k - i g B_0\frac{e}{2\omega_p} \tilde n_k=0.
\end{array}
\end{equation}
We now define dimensionless fields $$\mathcal{A}_k=\frac{e k}{2\omega_p m_e}\varphi_k, \quad \mathcal{B}_k =\frac{n_k}{n_0},$$ and replace $\omega=-i \partial/\partial t$ to obtain
\begin{equation}
\begin{array}{c}
\frac{\partial \mathcal{B}_k}{\partial t} = -i \omega_p \mathcal{B}_k+i g B_0 \mathcal{A}_k, \\\\
\frac{\partial \mathcal{A}_k}{\partial t} = -i\omega_\varphi \mathcal{A}_k-i \frac{g B_0}{4} \mathcal{B}_k
\end{array}
\label{eq_evolution1}
\end{equation}
\begin{figure}[t!]
\includegraphics[scale=0.6]{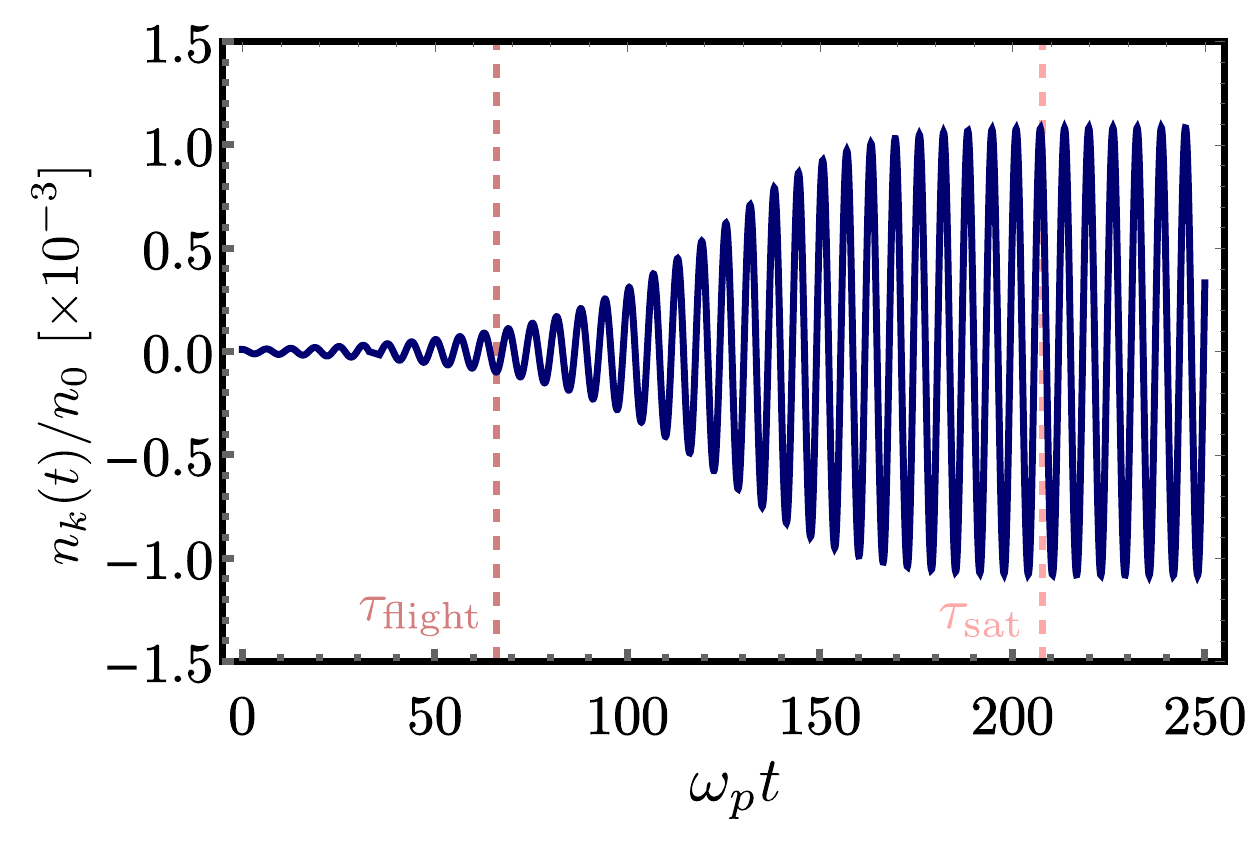}
\includegraphics[scale=0.6]{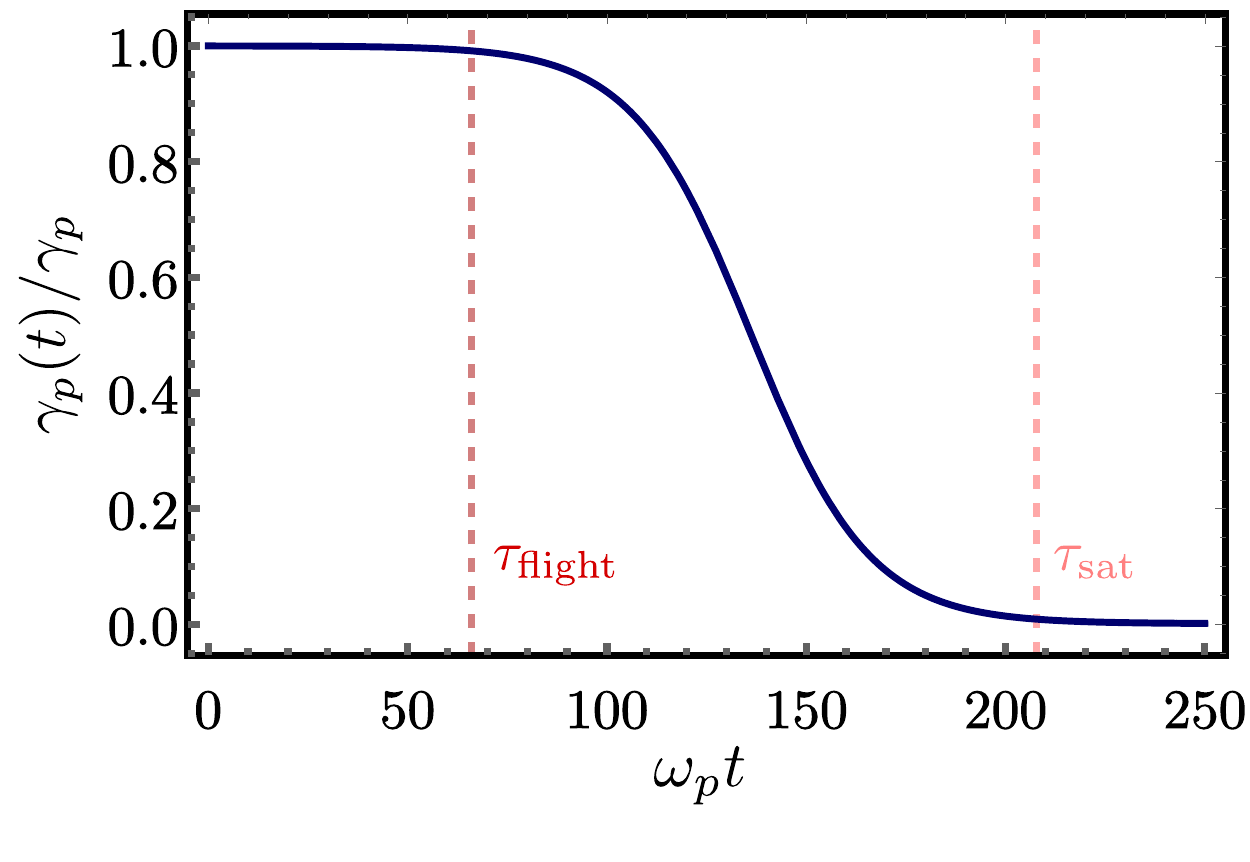}
\caption{Evolution of the beam-plasma instability for the most unstable mode, $k\simeq \omega_p/u_0$. The top panel depicts the electron density, undergoing exponential growth at early stages and followed by a saturation in the nonlinear stage $t \gamma_p\gg 1$. The bottom panel depicts the nonlinear growth rate $\tilde\gamma_p$ as a function of time, vanishing after the saturation. 
In both panels, the lighter dashed line represents the travel time for an axion of mass $m_\varphi=0.1 \omega_p$ in a plasma column of size $L=3.15$ m, while the darker dashed line marks the saturation time $\tau_{\rm sat}$. In the numerical calculations, we have taken $\omega_p=2\pi \times 10$ GHz, $g=10^{-13}$ GeV$^{-1}$, $B_0=1$ T, $f=0.1$ and $u_0=0.99$, considering the instability to start from thermal noise, $n_k(0)\sim 10^{-6}n_0$.}
\label{fig_supp_1}
\end{figure}
In the following, we show that this two-mode model is sufficient to understand the resonant mode conversion process taking place in the unstable plasma, avoiding the necessity to solve the cumbersome problem of introducing the beam dynamics. As argued in the main text, this is possible because the smallness of the axion-photon coupling in respect to the plasma frequency allows for the following separation of scales $$\omega_p\gg g B_0 \gg \nu g B_0.$$ As such, the beam-plasma and the plasma-axion problems can treated iteratively, with the effect of the beam in the axions being introduced {it a posteriori}. This is done by allowing the plasma frequency to become complex, $\omega_p\rightarrow \omega_r+i\gamma_p$, with $\omega_r=u_0k(1-\nu^{1/3})$, as discussed in the manuscript, and $\gamma_p$ being given by Eq. (12) of the main text. Solving the eigenvalue problem in Eq. \eqref{eq_evolution1} with the boundary conditions $\mathcal{B}_k(0)=1$ and $\mathcal{A}_k=0$, we can calculate the  axion-plasmon conversion probability of a certain mode $k$, $P_{p\rightarrow \varphi}(t)=\vert \langle \mathcal{A}_k(t)\vert \mathcal{B}_k(0)\rangle\vert^2$ to be 
\begin{equation}
P_{p\rightarrow \varphi}(t)=	\left\vert\frac{g^2B_0^2\sin^2\left[\frac{1}{2}t\sqrt{\left(\omega_r-\omega_\varphi+i\gamma_p\right)^2-g^2B_0^2}\right]}{4 \left[g^2 B_0^2-\left(\omega_r-\omega_\varphi +i\gamma_p\right)^2\right]}\right\vert.
\label{eq_prob_unstable}
\end{equation}
The probability of conversion in a stable plasma can be easily recovered by taking $\gamma_p=0$ and $\omega_r=\omega_p$, reading
\begin{equation}
P^{\rm stable}_{p\rightarrow \varphi}(t)=	\left\vert\frac{g^2B_0^2\sin^2\left[\frac{1}{2}t\sqrt{\left(\omega_p-\omega_\varphi\right)^2-g^2B_0^2}\right]}{4 \left[g^2 B_0^2-\left(\omega_p-\omega_\varphi \right)^2\right]}\right\vert,
\end{equation}
being formally very similar to the expression found for the axion-photon conversion problem \cite{raffelt_1988}. Thanks to the imaginary part $i \gamma_p$, the conversion probability can be quite large for sufficiently large times (see Figs. \ref{fig_supp_1} and \ref{fig_prob_k} for illustration). In particular, at resonance, $\omega_r=\omega_\varphi$, we obtain 
\begin{equation}
P_{p\rightarrow\varphi}(t)\simeq \frac{g^2 B_0^2}{4\gamma_p^2}e^{\gamma_p t},
\label{eq_prob_linear}
\end{equation}
where we have used the fact $\gamma_p\gg gB_0$. For a plasma column the of size $L$, the axion time-of-flight can be determined as
\begin{equation}
\tau_{\rm flight}=\frac{L}{v_\varphi}\simeq \frac{L k}{\sqrt{k^2+m_\varphi^2}},
\end{equation}
neglecting the small mass correction $gB_0$ introduced by the plasma. If the latter is sufficiently large, then the saturation of the instability has to be taken into account, and the exponentially growing probability given in Eq. \eqref{eq_prob_linear} may no longer be valid. To take this effect into account, we investigate the beam-plasma instability within the quasi-linear diffusion approximation, where the effect of the electron trapping due to the plasma waves is considered \cite{oneil_1971, sharma_1976}. 
In a first approximation, saturation effects can be cast by letting the growth rate to become a time-dependent function 
\begin{equation}
\gamma_p(t)\simeq \gamma_p\left(1-\frac{9}{8}\frac{\omega_p^4}{\gamma_p^4}\frac{n_k(t)^2}{n_0^2}\right).
\end{equation}
Then, we replace $\gamma_p$ by $\gamma_p(t)$ in Eqs. \eqref{eq_evolution1} to compute the electron dynamics. As depicted in Fig. \ref{fig_supp_1}, the electron density ceases to grow in the late stages of the evolution, and therefore the instability saturates after a time $\tau_{\rm sat}\sim \nu^{1/3}\gamma_p/\omega_p^2$. At this point, the growth rate vanishes. This is accompanied by a saturation of the conversion probability, as depicted in Fig. \ref{fig_prob_k}. For sufficiently light axions, however, $\tau_{\rm flight}<\tau_{\rm sat}$, and therefore Eq. \eqref{eq_prob_unstable} can be used in the evaluation of the probability. On the contrary, heavier axions may emerge deep in the nonlinear regime and, therefore, we make use the piecewise function discussed in the main text, 
\begin{equation}
P_{p\rightarrow \varphi}=\left\lbrace
\begin{array}{c}
e^{2\gamma_p t}P_{p\rightarrow \varphi}^{\rm osc}, 	\quad \tau_{\rm flight}\leq \tau_{\rm sat} \\\\
e^{2\gamma_p \tau_{\rm sat}}P_{p\rightarrow \varphi}^{\rm osc}, \quad \tau_{\rm flight}>\tau_{\rm sat} 
\end{array} 
\right. ,
\label{eq_prob_final}
\end{equation}
where 
\begin{equation}
\displaystyle{
P_{p\rightarrow \varphi}^{\rm osc}=\left\vert \frac{g^2B_0^2 \sin^2 \left[\frac{t}{2}\sqrt{g^2B_0^2-(\omega_r-\omega_\varphi)^2}\right]}{4\left[g^2B_0^2-(\omega_r-\omega_\varphi)^2-\gamma_p^2\right]} \right \vert}
\end{equation}
is the oscillating (non-growing) probability in the plasma. As it is patent from Fig. \ref{fig_prob_k}, this is an excellent interpolation between the linear and nonlinear stages.

\bibliographystyle{apsrev4-1}
\bibliography{references.bib}

\begin{thebibliography}{58}%
\makeatletter
\providecommand \@ifxundefined [1]{%
 \@ifx{#1\undefined}
}%
\providecommand \@ifnum [1]{%
 \ifnum #1\expandafter \@firstoftwo
 \else \expandafter \@secondoftwo
 \fi
}%
\providecommand \@ifx [1]{%
 \ifx #1\expandafter \@firstoftwo
 \else \expandafter \@secondoftwo
 \fi
}%
\providecommand \natexlab [1]{#1}%
\providecommand \enquote  [1]{``#1''}%
\providecommand \bibnamefont  [1]{#1}%
\providecommand \bibfnamefont [1]{#1}%
\providecommand \citenamefont [1]{#1}%
\providecommand \href@noop [0]{\@secondoftwo}%
\providecommand \href [0]{\begingroup \@sanitize@url \@href}%
\providecommand \@href[1]{\@@startlink{#1}\@@href}%
\providecommand \@@href[1]{\endgroup#1\@@endlink}%
\providecommand \@sanitize@url [0]{\catcode `\\12\catcode `\$12\catcode
  `\&12\catcode `\#12\catcode `\^12\catcode `\_12\catcode `\%12\relax}%
\providecommand \@@startlink[1]{}%
\providecommand \@@endlink[0]{}%
\providecommand \url  [0]{\begingroup\@sanitize@url \@url }%
\providecommand \@url [1]{\endgroup\@href {#1}{\urlprefix }}%
\providecommand \urlprefix  [0]{URL }%
\providecommand \Eprint [0]{\href }%
\providecommand \doibase [0]{http://dx.doi.org/}%
\providecommand \selectlanguage [0]{\@gobble}%
\providecommand \bibinfo  [0]{\@secondoftwo}%
\providecommand \bibfield  [0]{\@secondoftwo}%
\providecommand \translation [1]{[#1]}%
\providecommand \BibitemOpen [0]{}%
\providecommand \bibitemStop [0]{}%
\providecommand \bibitemNoStop [0]{.\EOS\space}%
\providecommand \EOS [0]{\spacefactor3000\relax}%
\providecommand \BibitemShut  [1]{\csname bibitem#1\endcsname}%
\let\auto@bib@innerbib\@empty
\bibitem [{\citenamefont {Pendlebury}\ \emph {et~al.}(2015)\citenamefont
  {Pendlebury}, \citenamefont {Afach}, \citenamefont {Ayres}, \citenamefont
  {Baker}, \citenamefont {Ban}, \citenamefont {Bison}, \citenamefont {Bodek},
  \citenamefont {Burghoff}, \citenamefont {Geltenbort}, \citenamefont {Green},
  \citenamefont {Griffith}, \citenamefont {van~der Grinten}, \citenamefont
  {Gruji\ifmmode~\acute{c}\else \'{c}\fi{}}, \citenamefont {Harris},
  \citenamefont {H\'elaine}, \citenamefont {Iaydjiev}, \citenamefont {Ivanov},
  \citenamefont {Kasprzak}, \citenamefont {Kermaidic}, \citenamefont {Kirch},
  \citenamefont {Koch}, \citenamefont {Komposch}, \citenamefont {Kozela},
  \citenamefont {Krempel}, \citenamefont {Lauss}, \citenamefont {Lefort},
  \citenamefont {Lemi\`ere}, \citenamefont {May}, \citenamefont {Musgrave},
  \citenamefont {Naviliat-Cuncic}, \citenamefont {Piegsa}, \citenamefont
  {Pignol}, \citenamefont {Prashanth}, \citenamefont {Qu\'em\'ener},
  \citenamefont {Rawlik}, \citenamefont {Rebreyend}, \citenamefont
  {Richardson}, \citenamefont {Ries}, \citenamefont {Roccia}, \citenamefont
  {Rozpedzik}, \citenamefont {Schnabel}, \citenamefont {Schmidt-Wellenburg},
  \citenamefont {Severijns}, \citenamefont {Shiers}, \citenamefont {Thorne},
  \citenamefont {Weis}, \citenamefont {Winston}, \citenamefont {Wursten},
  \citenamefont {Zejma},\ and\ \citenamefont {Zsigmond}}]{pendlebury_2015}%
  \BibitemOpen
  \bibfield  {author} {\bibinfo {author} {\bibfnamefont {J.~M.}\ \bibnamefont
  {Pendlebury}}, \bibinfo {author} {\bibfnamefont {S.}~\bibnamefont {Afach}},
  \bibinfo {author} {\bibfnamefont {N.~J.}\ \bibnamefont {Ayres}}, \bibinfo
  {author} {\bibfnamefont {C.~A.}\ \bibnamefont {Baker}}, \bibinfo {author}
  {\bibfnamefont {G.}~\bibnamefont {Ban}}, \bibinfo {author} {\bibfnamefont
  {G.}~\bibnamefont {Bison}}, \bibinfo {author} {\bibfnamefont
  {K.}~\bibnamefont {Bodek}}, \bibinfo {author} {\bibfnamefont
  {M.}~\bibnamefont {Burghoff}}, \bibinfo {author} {\bibfnamefont
  {P.}~\bibnamefont {Geltenbort}}, \bibinfo {author} {\bibfnamefont
  {K.}~\bibnamefont {Green}}, \bibinfo {author} {\bibfnamefont {W.~C.}\
  \bibnamefont {Griffith}}, \bibinfo {author} {\bibfnamefont {M.}~\bibnamefont
  {van~der Grinten}}, \bibinfo {author} {\bibfnamefont {Z.~D.}\ \bibnamefont
  {Gruji\ifmmode~\acute{c}\else \'{c}\fi{}}}, \bibinfo {author} {\bibfnamefont
  {P.~G.}\ \bibnamefont {Harris}}, \bibinfo {author} {\bibfnamefont
  {V.}~\bibnamefont {H\'elaine}}, \bibinfo {author} {\bibfnamefont
  {P.}~\bibnamefont {Iaydjiev}}, \bibinfo {author} {\bibfnamefont {S.~N.}\
  \bibnamefont {Ivanov}}, \bibinfo {author} {\bibfnamefont {M.}~\bibnamefont
  {Kasprzak}}, \bibinfo {author} {\bibfnamefont {Y.}~\bibnamefont {Kermaidic}},
  \bibinfo {author} {\bibfnamefont {K.}~\bibnamefont {Kirch}}, \bibinfo
  {author} {\bibfnamefont {H.-C.}\ \bibnamefont {Koch}}, \bibinfo {author}
  {\bibfnamefont {S.}~\bibnamefont {Komposch}}, \bibinfo {author}
  {\bibfnamefont {A.}~\bibnamefont {Kozela}}, \bibinfo {author} {\bibfnamefont
  {J.}~\bibnamefont {Krempel}}, \bibinfo {author} {\bibfnamefont
  {B.}~\bibnamefont {Lauss}}, \bibinfo {author} {\bibfnamefont
  {T.}~\bibnamefont {Lefort}}, \bibinfo {author} {\bibfnamefont
  {Y.}~\bibnamefont {Lemi\`ere}}, \bibinfo {author} {\bibfnamefont {D.~J.~R.}\
  \bibnamefont {May}}, \bibinfo {author} {\bibfnamefont {M.}~\bibnamefont
  {Musgrave}}, \bibinfo {author} {\bibfnamefont {O.}~\bibnamefont
  {Naviliat-Cuncic}}, \bibinfo {author} {\bibfnamefont {F.~M.}\ \bibnamefont
  {Piegsa}}, \bibinfo {author} {\bibfnamefont {G.}~\bibnamefont {Pignol}},
  \bibinfo {author} {\bibfnamefont {P.~N.}\ \bibnamefont {Prashanth}}, \bibinfo
  {author} {\bibfnamefont {G.}~\bibnamefont {Qu\'em\'ener}}, \bibinfo {author}
  {\bibfnamefont {M.}~\bibnamefont {Rawlik}}, \bibinfo {author} {\bibfnamefont
  {D.}~\bibnamefont {Rebreyend}}, \bibinfo {author} {\bibfnamefont {J.~D.}\
  \bibnamefont {Richardson}}, \bibinfo {author} {\bibfnamefont
  {D.}~\bibnamefont {Ries}}, \bibinfo {author} {\bibfnamefont {S.}~\bibnamefont
  {Roccia}}, \bibinfo {author} {\bibfnamefont {D.}~\bibnamefont {Rozpedzik}},
  \bibinfo {author} {\bibfnamefont {A.}~\bibnamefont {Schnabel}}, \bibinfo
  {author} {\bibfnamefont {P.}~\bibnamefont {Schmidt-Wellenburg}}, \bibinfo
  {author} {\bibfnamefont {N.}~\bibnamefont {Severijns}}, \bibinfo {author}
  {\bibfnamefont {D.}~\bibnamefont {Shiers}}, \bibinfo {author} {\bibfnamefont
  {J.~A.}\ \bibnamefont {Thorne}}, \bibinfo {author} {\bibfnamefont
  {A.}~\bibnamefont {Weis}}, \bibinfo {author} {\bibfnamefont {O.~J.}\
  \bibnamefont {Winston}}, \bibinfo {author} {\bibfnamefont {E.}~\bibnamefont
  {Wursten}}, \bibinfo {author} {\bibfnamefont {J.}~\bibnamefont {Zejma}}, \
  and\ \bibinfo {author} {\bibfnamefont {G.}~\bibnamefont {Zsigmond}},\ }\href
  {\doibase 10.1103/PhysRevD.92.092003} {\bibfield  {journal} {\bibinfo
  {journal} {Phys. Rev. D}\ }\textbf {\bibinfo {volume} {92}},\ \bibinfo
  {pages} {092003} (\bibinfo {year} {2015})}\BibitemShut {NoStop}%
\bibitem [{\citenamefont {Jaeckel}\ and\ \citenamefont
  {Ringwald}(2010)}]{jaeckel_2010}%
  \BibitemOpen
  \bibfield  {author} {\bibinfo {author} {\bibfnamefont {J.}~\bibnamefont
  {Jaeckel}}\ and\ \bibinfo {author} {\bibfnamefont {A.}~\bibnamefont
  {Ringwald}},\ }\href {\doibase 10.1146/annurev.nucl.012809.104433} {\bibfield
   {journal} {\bibinfo  {journal} {Annual Review of Nuclear and Particle
  Science}\ }\textbf {\bibinfo {volume} {60}},\ \bibinfo {pages} {405}
  (\bibinfo {year} {2010})},\ \Eprint
  {http://arxiv.org/abs/https://doi.org/10.1146/annurev.nucl.012809.104433}
  {https://doi.org/10.1146/annurev.nucl.012809.104433} \BibitemShut {NoStop}%
\bibitem [{\citenamefont {Ringwald}(2012)}]{ringwald_2012}%
  \BibitemOpen
  \bibfield  {author} {\bibinfo {author} {\bibfnamefont {A.}~\bibnamefont
  {Ringwald}},\ }\href {\doibase https://doi.org/10.1016/j.dark.2012.10.008}
  {\bibfield  {journal} {\bibinfo  {journal} {Physics of the Dark Universe}\
  }\textbf {\bibinfo {volume} {1}},\ \bibinfo {pages} {116 } (\bibinfo {year}
  {2012})},\ \bibinfo {note} {next Decade in Dark Matter and Dark
  Energy}\BibitemShut {NoStop}%
\bibitem [{\citenamefont {Kim}\ and\ \citenamefont {Carosi}(2010)}]{kim_2010}%
  \BibitemOpen
  \bibfield  {author} {\bibinfo {author} {\bibfnamefont {J.~E.}\ \bibnamefont
  {Kim}}\ and\ \bibinfo {author} {\bibfnamefont {G.}~\bibnamefont {Carosi}},\
  }\href {\doibase 10.1103/RevModPhys.82.557} {\bibfield  {journal} {\bibinfo
  {journal} {Rev. Mod. Phys.}\ }\textbf {\bibinfo {volume} {82}},\ \bibinfo
  {pages} {557} (\bibinfo {year} {2010})}\BibitemShut {NoStop}%
\bibitem [{\citenamefont {Peccei}\ and\ \citenamefont
  {Quinn}(1977)}]{quinn_1977}%
  \BibitemOpen
  \bibfield  {author} {\bibinfo {author} {\bibfnamefont {R.~D.}\ \bibnamefont
  {Peccei}}\ and\ \bibinfo {author} {\bibfnamefont {H.~R.}\ \bibnamefont
  {Quinn}},\ }\href {\doibase 10.1103/PhysRevLett.38.1440} {\bibfield
  {journal} {\bibinfo  {journal} {Phys. Rev. Lett.}\ }\textbf {\bibinfo
  {volume} {38}},\ \bibinfo {pages} {1440} (\bibinfo {year}
  {1977})}\BibitemShut {NoStop}%
\bibitem [{\citenamefont {Weinberg}(1978)}]{weinberg_1978}%
  \BibitemOpen
  \bibfield  {author} {\bibinfo {author} {\bibfnamefont {S.}~\bibnamefont
  {Weinberg}},\ }\href {\doibase 10.1103/PhysRevLett.40.223} {\bibfield
  {journal} {\bibinfo  {journal} {Phys. Rev. Lett.}\ }\textbf {\bibinfo
  {volume} {40}},\ \bibinfo {pages} {223} (\bibinfo {year} {1978})}\BibitemShut
  {NoStop}%
\bibitem [{\citenamefont {Wilczek}(1978)}]{wilczek_1978}%
  \BibitemOpen
  \bibfield  {author} {\bibinfo {author} {\bibfnamefont {F.}~\bibnamefont
  {Wilczek}},\ }\href {\doibase 10.1103/PhysRevLett.40.279} {\bibfield
  {journal} {\bibinfo  {journal} {Phys. Rev. Lett.}\ }\textbf {\bibinfo
  {volume} {40}},\ \bibinfo {pages} {279} (\bibinfo {year} {1978})}\BibitemShut
  {NoStop}%
\bibitem [{\citenamefont {Sikivie}(1983)}]{PhysRevLett.51.1415}%
  \BibitemOpen
  \bibfield  {author} {\bibinfo {author} {\bibfnamefont {P.}~\bibnamefont
  {Sikivie}},\ }\href {\doibase 10.1103/PhysRevLett.51.1415} {\bibfield
  {journal} {\bibinfo  {journal} {Phys. Rev. Lett.}\ }\textbf {\bibinfo
  {volume} {51}},\ \bibinfo {pages} {1415} (\bibinfo {year}
  {1983})}\BibitemShut {NoStop}%
\bibitem [{\citenamefont {Asztalos}\ \emph
  {et~al.}(2010{\natexlab{a}})\citenamefont {Asztalos}, \citenamefont {Carosi},
  \citenamefont {Hagmann}, \citenamefont {Kinion}, \citenamefont {van Bibber},
  \citenamefont {Hotz}, \citenamefont {Rosenberg}, \citenamefont {Rybka},
  \citenamefont {Hoskins}, \citenamefont {Hwang}, \citenamefont {Sikivie},
  \citenamefont {Tanner}, \citenamefont {Bradley},\ and\ \citenamefont
  {Clarke}}]{PhysRevLett.104.041301}%
  \BibitemOpen
  \bibfield  {author} {\bibinfo {author} {\bibfnamefont {S.~J.}\ \bibnamefont
  {Asztalos}}, \bibinfo {author} {\bibfnamefont {G.}~\bibnamefont {Carosi}},
  \bibinfo {author} {\bibfnamefont {C.}~\bibnamefont {Hagmann}}, \bibinfo
  {author} {\bibfnamefont {D.}~\bibnamefont {Kinion}}, \bibinfo {author}
  {\bibfnamefont {K.}~\bibnamefont {van Bibber}}, \bibinfo {author}
  {\bibfnamefont {M.}~\bibnamefont {Hotz}}, \bibinfo {author} {\bibfnamefont
  {L.~J.}\ \bibnamefont {Rosenberg}}, \bibinfo {author} {\bibfnamefont
  {G.}~\bibnamefont {Rybka}}, \bibinfo {author} {\bibfnamefont
  {J.}~\bibnamefont {Hoskins}}, \bibinfo {author} {\bibfnamefont
  {J.}~\bibnamefont {Hwang}}, \bibinfo {author} {\bibfnamefont
  {P.}~\bibnamefont {Sikivie}}, \bibinfo {author} {\bibfnamefont {D.~B.}\
  \bibnamefont {Tanner}}, \bibinfo {author} {\bibfnamefont {R.}~\bibnamefont
  {Bradley}}, \ and\ \bibinfo {author} {\bibfnamefont {J.}~\bibnamefont
  {Clarke}},\ }\href {\doibase 10.1103/PhysRevLett.104.041301} {\bibfield
  {journal} {\bibinfo  {journal} {Phys. Rev. Lett.}\ }\textbf {\bibinfo
  {volume} {104}},\ \bibinfo {pages} {041301} (\bibinfo {year}
  {2010}{\natexlab{a}})}\BibitemShut {NoStop}%
\bibitem [{\citenamefont {Zioutas}\ \emph {et~al.}(2005)\citenamefont
  {Zioutas}, \citenamefont {Andriamonje}, \citenamefont {Arsov}, \citenamefont
  {Aune}, \citenamefont {Autiero}, \citenamefont {Avignone}, \citenamefont
  {Barth}, \citenamefont {Belov}, \citenamefont {Beltr\'an}, \citenamefont
  {Br\"auninger}, \citenamefont {Carmona}, \citenamefont {Cebri\'an},
  \citenamefont {Chesi}, \citenamefont {Collar}, \citenamefont {Creswick},
  \citenamefont {Dafni}, \citenamefont {Davenport}, \citenamefont {Di~Lella},
  \citenamefont {Eleftheriadis}, \citenamefont {Englhauser}, \citenamefont
  {Fanourakis}, \citenamefont {Farach}, \citenamefont {Ferrer}, \citenamefont
  {Fischer}, \citenamefont {Franz}, \citenamefont {Friedrich}, \citenamefont
  {Geralis}, \citenamefont {Giomataris}, \citenamefont {Gninenko},
  \citenamefont {Goloubev}, \citenamefont {Hasinoff}, \citenamefont {Heinsius},
  \citenamefont {Hoffmann}, \citenamefont {Irastorza}, \citenamefont {Jacoby},
  \citenamefont {Kang}, \citenamefont {K\"onigsmann}, \citenamefont {Kotthaus},
  \citenamefont {Kr\ifmmode~\check{c}\else \v{c}\fi{}mar}, \citenamefont
  {Kousouris}, \citenamefont {Kuster}, \citenamefont
  {Laki\ifmmode~\acute{c}\else \'{c}\fi{}}, \citenamefont {Lasseur},
  \citenamefont {Liolios}, \citenamefont {Ljubi\ifmmode \check{c}\else
  \v{c}\fi{}i\ifmmode~\acute{c}\else \'{c}\fi{}}, \citenamefont {Lutz},
  \citenamefont {Luz\'on}, \citenamefont {Miller}, \citenamefont {Morales},
  \citenamefont {Morales}, \citenamefont {Mutterer}, \citenamefont
  {Nikolaidis}, \citenamefont {Ortiz}, \citenamefont {Papaevangelou},
  \citenamefont {Placci}, \citenamefont {Raffelt}, \citenamefont {Ruz},
  \citenamefont {Riege}, \citenamefont {Sarsa}, \citenamefont {Savvidis},
  \citenamefont {Serber}, \citenamefont {Serpico}, \citenamefont {Semertzidis},
  \citenamefont {Stewart}, \citenamefont {Vieira}, \citenamefont {Villar},
  \citenamefont {Walckiers},\ and\ \citenamefont
  {Zachariadou}}]{PhysRevLett.94.121301}%
  \BibitemOpen
  \bibfield  {author} {\bibinfo {author} {\bibfnamefont {K.}~\bibnamefont
  {Zioutas}}, \bibinfo {author} {\bibfnamefont {S.}~\bibnamefont
  {Andriamonje}}, \bibinfo {author} {\bibfnamefont {V.}~\bibnamefont {Arsov}},
  \bibinfo {author} {\bibfnamefont {S.}~\bibnamefont {Aune}}, \bibinfo {author}
  {\bibfnamefont {D.}~\bibnamefont {Autiero}}, \bibinfo {author} {\bibfnamefont
  {F.~T.}\ \bibnamefont {Avignone}}, \bibinfo {author} {\bibfnamefont
  {K.}~\bibnamefont {Barth}}, \bibinfo {author} {\bibfnamefont
  {A.}~\bibnamefont {Belov}}, \bibinfo {author} {\bibfnamefont
  {B.}~\bibnamefont {Beltr\'an}}, \bibinfo {author} {\bibfnamefont
  {H.}~\bibnamefont {Br\"auninger}}, \bibinfo {author} {\bibfnamefont {J.~M.}\
  \bibnamefont {Carmona}}, \bibinfo {author} {\bibfnamefont {S.}~\bibnamefont
  {Cebri\'an}}, \bibinfo {author} {\bibfnamefont {E.}~\bibnamefont {Chesi}},
  \bibinfo {author} {\bibfnamefont {J.~I.}\ \bibnamefont {Collar}}, \bibinfo
  {author} {\bibfnamefont {R.}~\bibnamefont {Creswick}}, \bibinfo {author}
  {\bibfnamefont {T.}~\bibnamefont {Dafni}}, \bibinfo {author} {\bibfnamefont
  {M.}~\bibnamefont {Davenport}}, \bibinfo {author} {\bibfnamefont
  {L.}~\bibnamefont {Di~Lella}}, \bibinfo {author} {\bibfnamefont
  {C.}~\bibnamefont {Eleftheriadis}}, \bibinfo {author} {\bibfnamefont
  {J.}~\bibnamefont {Englhauser}}, \bibinfo {author} {\bibfnamefont
  {G.}~\bibnamefont {Fanourakis}}, \bibinfo {author} {\bibfnamefont
  {H.}~\bibnamefont {Farach}}, \bibinfo {author} {\bibfnamefont
  {E.}~\bibnamefont {Ferrer}}, \bibinfo {author} {\bibfnamefont
  {H.}~\bibnamefont {Fischer}}, \bibinfo {author} {\bibfnamefont
  {J.}~\bibnamefont {Franz}}, \bibinfo {author} {\bibfnamefont
  {P.}~\bibnamefont {Friedrich}}, \bibinfo {author} {\bibfnamefont
  {T.}~\bibnamefont {Geralis}}, \bibinfo {author} {\bibfnamefont
  {I.}~\bibnamefont {Giomataris}}, \bibinfo {author} {\bibfnamefont
  {S.}~\bibnamefont {Gninenko}}, \bibinfo {author} {\bibfnamefont
  {N.}~\bibnamefont {Goloubev}}, \bibinfo {author} {\bibfnamefont {M.~D.}\
  \bibnamefont {Hasinoff}}, \bibinfo {author} {\bibfnamefont {F.~H.}\
  \bibnamefont {Heinsius}}, \bibinfo {author} {\bibfnamefont {D.~H.~H.}\
  \bibnamefont {Hoffmann}}, \bibinfo {author} {\bibfnamefont {I.~G.}\
  \bibnamefont {Irastorza}}, \bibinfo {author} {\bibfnamefont {J.}~\bibnamefont
  {Jacoby}}, \bibinfo {author} {\bibfnamefont {D.}~\bibnamefont {Kang}},
  \bibinfo {author} {\bibfnamefont {K.}~\bibnamefont {K\"onigsmann}}, \bibinfo
  {author} {\bibfnamefont {R.}~\bibnamefont {Kotthaus}}, \bibinfo {author}
  {\bibfnamefont {M.}~\bibnamefont {Kr\ifmmode~\check{c}\else \v{c}\fi{}mar}},
  \bibinfo {author} {\bibfnamefont {K.}~\bibnamefont {Kousouris}}, \bibinfo
  {author} {\bibfnamefont {M.}~\bibnamefont {Kuster}}, \bibinfo {author}
  {\bibfnamefont {B.}~\bibnamefont {Laki\ifmmode~\acute{c}\else \'{c}\fi{}}},
  \bibinfo {author} {\bibfnamefont {C.}~\bibnamefont {Lasseur}}, \bibinfo
  {author} {\bibfnamefont {A.}~\bibnamefont {Liolios}}, \bibinfo {author}
  {\bibfnamefont {A.}~\bibnamefont {Ljubi\ifmmode \check{c}\else
  \v{c}\fi{}i\ifmmode~\acute{c}\else \'{c}\fi{}}}, \bibinfo {author}
  {\bibfnamefont {G.}~\bibnamefont {Lutz}}, \bibinfo {author} {\bibfnamefont
  {G.}~\bibnamefont {Luz\'on}}, \bibinfo {author} {\bibfnamefont {D.~W.}\
  \bibnamefont {Miller}}, \bibinfo {author} {\bibfnamefont {A.}~\bibnamefont
  {Morales}}, \bibinfo {author} {\bibfnamefont {J.}~\bibnamefont {Morales}},
  \bibinfo {author} {\bibfnamefont {M.}~\bibnamefont {Mutterer}}, \bibinfo
  {author} {\bibfnamefont {A.}~\bibnamefont {Nikolaidis}}, \bibinfo {author}
  {\bibfnamefont {A.}~\bibnamefont {Ortiz}}, \bibinfo {author} {\bibfnamefont
  {T.}~\bibnamefont {Papaevangelou}}, \bibinfo {author} {\bibfnamefont
  {A.}~\bibnamefont {Placci}}, \bibinfo {author} {\bibfnamefont
  {G.}~\bibnamefont {Raffelt}}, \bibinfo {author} {\bibfnamefont
  {J.}~\bibnamefont {Ruz}}, \bibinfo {author} {\bibfnamefont {H.}~\bibnamefont
  {Riege}}, \bibinfo {author} {\bibfnamefont {M.~L.}\ \bibnamefont {Sarsa}},
  \bibinfo {author} {\bibfnamefont {I.}~\bibnamefont {Savvidis}}, \bibinfo
  {author} {\bibfnamefont {W.}~\bibnamefont {Serber}}, \bibinfo {author}
  {\bibfnamefont {P.}~\bibnamefont {Serpico}}, \bibinfo {author} {\bibfnamefont
  {Y.}~\bibnamefont {Semertzidis}}, \bibinfo {author} {\bibfnamefont
  {L.}~\bibnamefont {Stewart}}, \bibinfo {author} {\bibfnamefont {J.~D.}\
  \bibnamefont {Vieira}}, \bibinfo {author} {\bibfnamefont {J.}~\bibnamefont
  {Villar}}, \bibinfo {author} {\bibfnamefont {L.}~\bibnamefont {Walckiers}}, \
  and\ \bibinfo {author} {\bibfnamefont {K.}~\bibnamefont {Zachariadou}}
  (\bibinfo {collaboration} {CAST Collaboration}),\ }\href {\doibase
  10.1103/PhysRevLett.94.121301} {\bibfield  {journal} {\bibinfo  {journal}
  {Phys. Rev. Lett.}\ }\textbf {\bibinfo {volume} {94}},\ \bibinfo {pages}
  {121301} (\bibinfo {year} {2005})}\BibitemShut {NoStop}%
\bibitem [{\citenamefont {Fairbairn}\ \emph {et~al.}(2007)\citenamefont
  {Fairbairn}, \citenamefont {Rashba},\ and\ \citenamefont
  {Troitsky}}]{PhysRevLett.98.201801}%
  \BibitemOpen
  \bibfield  {author} {\bibinfo {author} {\bibfnamefont {M.}~\bibnamefont
  {Fairbairn}}, \bibinfo {author} {\bibfnamefont {T.}~\bibnamefont {Rashba}}, \
  and\ \bibinfo {author} {\bibfnamefont {S.}~\bibnamefont {Troitsky}},\ }\href
  {\doibase 10.1103/PhysRevLett.98.201801} {\bibfield  {journal} {\bibinfo
  {journal} {Phys. Rev. Lett.}\ }\textbf {\bibinfo {volume} {98}},\ \bibinfo
  {pages} {201801} (\bibinfo {year} {2007})}\BibitemShut {NoStop}%
\bibitem [{\citenamefont {Akerib}\ \emph {et~al.}(2017)\citenamefont {Akerib},
  \citenamefont {Alsum}, \citenamefont {Aquino}, \citenamefont {Ara\'ujo},
  \citenamefont {Bai}, \citenamefont {Bailey}, \citenamefont {Balajthy},
  \citenamefont {Beltrame}, \citenamefont {Bernard}, \citenamefont {Bernstein},
  \citenamefont {Biesiadzinski}, \citenamefont {Boulton}, \citenamefont
  {Br\'as}, \citenamefont {Byram}, \citenamefont {Cahn}, \citenamefont
  {Carmona-Benitez}, \citenamefont {Chan}, \citenamefont {Chiller},
  \citenamefont {Chiller}, \citenamefont {Currie}, \citenamefont {Cutter},
  \citenamefont {Davison}, \citenamefont {Dobi}, \citenamefont {Dobson},
  \citenamefont {Druszkiewicz}, \citenamefont {Edwards}, \citenamefont {Faham},
  \citenamefont {Fallon}, \citenamefont {Fiorucci}, \citenamefont {Gaitskell},
  \citenamefont {Gehman}, \citenamefont {Ghag}, \citenamefont {Gibson},
  \citenamefont {Gilchriese}, \citenamefont {Hall}, \citenamefont {Hanhardt},
  \citenamefont {Haselschwardt}, \citenamefont {Hertel}, \citenamefont {Hogan},
  \citenamefont {Horn}, \citenamefont {Huang}, \citenamefont {Ignarra},
  \citenamefont {Jacobsen}, \citenamefont {Ji}, \citenamefont {Kamdin},
  \citenamefont {Kazkaz}, \citenamefont {Khaitan}, \citenamefont {Knoche},
  \citenamefont {Larsen}, \citenamefont {Lee}, \citenamefont {Lenardo},
  \citenamefont {Lesko}, \citenamefont {Lindote}, \citenamefont {Lopes},
  \citenamefont {Manalaysay}, \citenamefont {Mannino}, \citenamefont
  {Marzioni}, \citenamefont {McKinsey}, \citenamefont {Mei}, \citenamefont
  {Mock}, \citenamefont {Moongweluwan}, \citenamefont {Morad}, \citenamefont
  {Murphy}, \citenamefont {Nehrkorn}, \citenamefont {Nelson}, \citenamefont
  {Neves}, \citenamefont {O'Sullivan}, \citenamefont {Oliver-Mallory},
  \citenamefont {Palladino}, \citenamefont {Pease}, \citenamefont {Reichhart},
  \citenamefont {Rhyne}, \citenamefont {Shaw}, \citenamefont {Shutt},
  \citenamefont {Silva}, \citenamefont {Solmaz}, \citenamefont {Solovov},
  \citenamefont {Sorensen}, \citenamefont {Stephenson}, \citenamefont {Sumner},
  \citenamefont {Szydagis}, \citenamefont {Taylor}, \citenamefont {Taylor},
  \citenamefont {Tennyson}, \citenamefont {Terman}, \citenamefont {Tiedt},
  \citenamefont {To}, \citenamefont {Tripathi}, \citenamefont {Tvrznikova},
  \citenamefont {Uvarov}, \citenamefont {Velan}, \citenamefont {Verbus},
  \citenamefont {Webb}, \citenamefont {White}, \citenamefont {Whitis},
  \citenamefont {Witherell}, \citenamefont {Wolfs}, \citenamefont {Xu},
  \citenamefont {Yazdani}, \citenamefont {Young},\ and\ \citenamefont
  {Zhang}}]{PhysRevLett.118.261301}%
  \BibitemOpen
  \bibfield  {author} {\bibinfo {author} {\bibfnamefont {D.~S.}\ \bibnamefont
  {Akerib}}, \bibinfo {author} {\bibfnamefont {S.}~\bibnamefont {Alsum}},
  \bibinfo {author} {\bibfnamefont {C.}~\bibnamefont {Aquino}}, \bibinfo
  {author} {\bibfnamefont {H.~M.}\ \bibnamefont {Ara\'ujo}}, \bibinfo {author}
  {\bibfnamefont {X.}~\bibnamefont {Bai}}, \bibinfo {author} {\bibfnamefont
  {A.~J.}\ \bibnamefont {Bailey}}, \bibinfo {author} {\bibfnamefont
  {J.}~\bibnamefont {Balajthy}}, \bibinfo {author} {\bibfnamefont
  {P.}~\bibnamefont {Beltrame}}, \bibinfo {author} {\bibfnamefont {E.~P.}\
  \bibnamefont {Bernard}}, \bibinfo {author} {\bibfnamefont {A.}~\bibnamefont
  {Bernstein}}, \bibinfo {author} {\bibfnamefont {T.~P.}\ \bibnamefont
  {Biesiadzinski}}, \bibinfo {author} {\bibfnamefont {E.~M.}\ \bibnamefont
  {Boulton}}, \bibinfo {author} {\bibfnamefont {P.}~\bibnamefont {Br\'as}},
  \bibinfo {author} {\bibfnamefont {D.}~\bibnamefont {Byram}}, \bibinfo
  {author} {\bibfnamefont {S.~B.}\ \bibnamefont {Cahn}}, \bibinfo {author}
  {\bibfnamefont {M.~C.}\ \bibnamefont {Carmona-Benitez}}, \bibinfo {author}
  {\bibfnamefont {C.}~\bibnamefont {Chan}}, \bibinfo {author} {\bibfnamefont
  {A.~A.}\ \bibnamefont {Chiller}}, \bibinfo {author} {\bibfnamefont
  {C.}~\bibnamefont {Chiller}}, \bibinfo {author} {\bibfnamefont
  {A.}~\bibnamefont {Currie}}, \bibinfo {author} {\bibfnamefont {J.~E.}\
  \bibnamefont {Cutter}}, \bibinfo {author} {\bibfnamefont {T.~J.~R.}\
  \bibnamefont {Davison}}, \bibinfo {author} {\bibfnamefont {A.}~\bibnamefont
  {Dobi}}, \bibinfo {author} {\bibfnamefont {J.~E.~Y.}\ \bibnamefont {Dobson}},
  \bibinfo {author} {\bibfnamefont {E.}~\bibnamefont {Druszkiewicz}}, \bibinfo
  {author} {\bibfnamefont {B.~N.}\ \bibnamefont {Edwards}}, \bibinfo {author}
  {\bibfnamefont {C.~H.}\ \bibnamefont {Faham}}, \bibinfo {author}
  {\bibfnamefont {S.~R.}\ \bibnamefont {Fallon}}, \bibinfo {author}
  {\bibfnamefont {S.}~\bibnamefont {Fiorucci}}, \bibinfo {author}
  {\bibfnamefont {R.~J.}\ \bibnamefont {Gaitskell}}, \bibinfo {author}
  {\bibfnamefont {V.~M.}\ \bibnamefont {Gehman}}, \bibinfo {author}
  {\bibfnamefont {C.}~\bibnamefont {Ghag}}, \bibinfo {author} {\bibfnamefont
  {K.~R.}\ \bibnamefont {Gibson}}, \bibinfo {author} {\bibfnamefont {M.~G.~D.}\
  \bibnamefont {Gilchriese}}, \bibinfo {author} {\bibfnamefont {C.~R.}\
  \bibnamefont {Hall}}, \bibinfo {author} {\bibfnamefont {M.}~\bibnamefont
  {Hanhardt}}, \bibinfo {author} {\bibfnamefont {S.~J.}\ \bibnamefont
  {Haselschwardt}}, \bibinfo {author} {\bibfnamefont {S.~A.}\ \bibnamefont
  {Hertel}}, \bibinfo {author} {\bibfnamefont {D.~P.}\ \bibnamefont {Hogan}},
  \bibinfo {author} {\bibfnamefont {M.}~\bibnamefont {Horn}}, \bibinfo {author}
  {\bibfnamefont {D.~Q.}\ \bibnamefont {Huang}}, \bibinfo {author}
  {\bibfnamefont {C.~M.}\ \bibnamefont {Ignarra}}, \bibinfo {author}
  {\bibfnamefont {R.~G.}\ \bibnamefont {Jacobsen}}, \bibinfo {author}
  {\bibfnamefont {W.}~\bibnamefont {Ji}}, \bibinfo {author} {\bibfnamefont
  {K.}~\bibnamefont {Kamdin}}, \bibinfo {author} {\bibfnamefont
  {K.}~\bibnamefont {Kazkaz}}, \bibinfo {author} {\bibfnamefont
  {D.}~\bibnamefont {Khaitan}}, \bibinfo {author} {\bibfnamefont
  {R.}~\bibnamefont {Knoche}}, \bibinfo {author} {\bibfnamefont {N.~A.}\
  \bibnamefont {Larsen}}, \bibinfo {author} {\bibfnamefont {C.}~\bibnamefont
  {Lee}}, \bibinfo {author} {\bibfnamefont {B.~G.}\ \bibnamefont {Lenardo}},
  \bibinfo {author} {\bibfnamefont {K.~T.}\ \bibnamefont {Lesko}}, \bibinfo
  {author} {\bibfnamefont {A.}~\bibnamefont {Lindote}}, \bibinfo {author}
  {\bibfnamefont {M.~I.}\ \bibnamefont {Lopes}}, \bibinfo {author}
  {\bibfnamefont {A.}~\bibnamefont {Manalaysay}}, \bibinfo {author}
  {\bibfnamefont {R.~L.}\ \bibnamefont {Mannino}}, \bibinfo {author}
  {\bibfnamefont {M.~F.}\ \bibnamefont {Marzioni}}, \bibinfo {author}
  {\bibfnamefont {D.~N.}\ \bibnamefont {McKinsey}}, \bibinfo {author}
  {\bibfnamefont {D.-M.}\ \bibnamefont {Mei}}, \bibinfo {author} {\bibfnamefont
  {J.}~\bibnamefont {Mock}}, \bibinfo {author} {\bibfnamefont {M.}~\bibnamefont
  {Moongweluwan}}, \bibinfo {author} {\bibfnamefont {J.~A.}\ \bibnamefont
  {Morad}}, \bibinfo {author} {\bibfnamefont {A.~S.~J.}\ \bibnamefont
  {Murphy}}, \bibinfo {author} {\bibfnamefont {C.}~\bibnamefont {Nehrkorn}},
  \bibinfo {author} {\bibfnamefont {H.~N.}\ \bibnamefont {Nelson}}, \bibinfo
  {author} {\bibfnamefont {F.}~\bibnamefont {Neves}}, \bibinfo {author}
  {\bibfnamefont {K.}~\bibnamefont {O'Sullivan}}, \bibinfo {author}
  {\bibfnamefont {K.~C.}\ \bibnamefont {Oliver-Mallory}}, \bibinfo {author}
  {\bibfnamefont {K.~J.}\ \bibnamefont {Palladino}}, \bibinfo {author}
  {\bibfnamefont {E.~K.}\ \bibnamefont {Pease}}, \bibinfo {author}
  {\bibfnamefont {L.}~\bibnamefont {Reichhart}}, \bibinfo {author}
  {\bibfnamefont {C.}~\bibnamefont {Rhyne}}, \bibinfo {author} {\bibfnamefont
  {S.}~\bibnamefont {Shaw}}, \bibinfo {author} {\bibfnamefont {T.~A.}\
  \bibnamefont {Shutt}}, \bibinfo {author} {\bibfnamefont {C.}~\bibnamefont
  {Silva}}, \bibinfo {author} {\bibfnamefont {M.}~\bibnamefont {Solmaz}},
  \bibinfo {author} {\bibfnamefont {V.~N.}\ \bibnamefont {Solovov}}, \bibinfo
  {author} {\bibfnamefont {P.}~\bibnamefont {Sorensen}}, \bibinfo {author}
  {\bibfnamefont {S.}~\bibnamefont {Stephenson}}, \bibinfo {author}
  {\bibfnamefont {T.~J.}\ \bibnamefont {Sumner}}, \bibinfo {author}
  {\bibfnamefont {M.}~\bibnamefont {Szydagis}}, \bibinfo {author}
  {\bibfnamefont {D.~J.}\ \bibnamefont {Taylor}}, \bibinfo {author}
  {\bibfnamefont {W.~C.}\ \bibnamefont {Taylor}}, \bibinfo {author}
  {\bibfnamefont {B.~P.}\ \bibnamefont {Tennyson}}, \bibinfo {author}
  {\bibfnamefont {P.~A.}\ \bibnamefont {Terman}}, \bibinfo {author}
  {\bibfnamefont {D.~R.}\ \bibnamefont {Tiedt}}, \bibinfo {author}
  {\bibfnamefont {W.~H.}\ \bibnamefont {To}}, \bibinfo {author} {\bibfnamefont
  {M.}~\bibnamefont {Tripathi}}, \bibinfo {author} {\bibfnamefont
  {L.}~\bibnamefont {Tvrznikova}}, \bibinfo {author} {\bibfnamefont
  {S.}~\bibnamefont {Uvarov}}, \bibinfo {author} {\bibfnamefont
  {V.}~\bibnamefont {Velan}}, \bibinfo {author} {\bibfnamefont {J.~R.}\
  \bibnamefont {Verbus}}, \bibinfo {author} {\bibfnamefont {R.~C.}\
  \bibnamefont {Webb}}, \bibinfo {author} {\bibfnamefont {J.~T.}\ \bibnamefont
  {White}}, \bibinfo {author} {\bibfnamefont {T.~J.}\ \bibnamefont {Whitis}},
  \bibinfo {author} {\bibfnamefont {M.~S.}\ \bibnamefont {Witherell}}, \bibinfo
  {author} {\bibfnamefont {F.~L.~H.}\ \bibnamefont {Wolfs}}, \bibinfo {author}
  {\bibfnamefont {J.}~\bibnamefont {Xu}}, \bibinfo {author} {\bibfnamefont
  {K.}~\bibnamefont {Yazdani}}, \bibinfo {author} {\bibfnamefont {S.~K.}\
  \bibnamefont {Young}}, \ and\ \bibinfo {author} {\bibfnamefont
  {C.}~\bibnamefont {Zhang}} (\bibinfo {collaboration} {LUX Collaboration}),\
  }\href {\doibase 10.1103/PhysRevLett.118.261301} {\bibfield  {journal}
  {\bibinfo  {journal} {Phys. Rev. Lett.}\ }\textbf {\bibinfo {volume} {118}},\
  \bibinfo {pages} {261301} (\bibinfo {year} {2017})}\BibitemShut {NoStop}%
\bibitem [{\citenamefont {Collaboration}(2017)}]{Collaboration2017}%
  \BibitemOpen
  \bibfield  {author} {\bibinfo {author} {\bibfnamefont {C.~A. S.~T.}\
  \bibnamefont {Collaboration}},\ }\href {http://dx.doi.org/10.1038/nphys4109}
  {\bibfield  {journal} {\bibinfo  {journal} {Nature Physics}\ }\textbf
  {\bibinfo {volume} {13}},\ \bibinfo {pages} {584 EP } (\bibinfo {year}
  {2017})},\ \bibinfo {note} {article}\BibitemShut {NoStop}%
\bibitem [{\citenamefont {Asztalos}\ \emph {et~al.}(2001)\citenamefont
  {Asztalos}, \citenamefont {Daw}, \citenamefont {Peng}, \citenamefont
  {Rosenberg}, \citenamefont {Hagmann}, \citenamefont {Kinion}, \citenamefont
  {Stoeffl}, \citenamefont {van Bibber}, \citenamefont {Sikivie}, \citenamefont
  {Sullivan}, \citenamefont {Tanner}, \citenamefont {Nezrick}, \citenamefont
  {Turner}, \citenamefont {Moltz}, \citenamefont {Powell}, \citenamefont
  {Andr\'e}, \citenamefont {Clarke}, \citenamefont {M\"uck},\ and\
  \citenamefont {Bradley}}]{asztalos_2001}%
  \BibitemOpen
  \bibfield  {author} {\bibinfo {author} {\bibfnamefont {S.}~\bibnamefont
  {Asztalos}}, \bibinfo {author} {\bibfnamefont {E.}~\bibnamefont {Daw}},
  \bibinfo {author} {\bibfnamefont {H.}~\bibnamefont {Peng}}, \bibinfo {author}
  {\bibfnamefont {L.~J.}\ \bibnamefont {Rosenberg}}, \bibinfo {author}
  {\bibfnamefont {C.}~\bibnamefont {Hagmann}}, \bibinfo {author} {\bibfnamefont
  {D.}~\bibnamefont {Kinion}}, \bibinfo {author} {\bibfnamefont
  {W.}~\bibnamefont {Stoeffl}}, \bibinfo {author} {\bibfnamefont
  {K.}~\bibnamefont {van Bibber}}, \bibinfo {author} {\bibfnamefont
  {P.}~\bibnamefont {Sikivie}}, \bibinfo {author} {\bibfnamefont {N.~S.}\
  \bibnamefont {Sullivan}}, \bibinfo {author} {\bibfnamefont {D.~B.}\
  \bibnamefont {Tanner}}, \bibinfo {author} {\bibfnamefont {F.}~\bibnamefont
  {Nezrick}}, \bibinfo {author} {\bibfnamefont {M.~S.}\ \bibnamefont {Turner}},
  \bibinfo {author} {\bibfnamefont {D.~M.}\ \bibnamefont {Moltz}}, \bibinfo
  {author} {\bibfnamefont {J.}~\bibnamefont {Powell}}, \bibinfo {author}
  {\bibfnamefont {M.-O.}\ \bibnamefont {Andr\'e}}, \bibinfo {author}
  {\bibfnamefont {J.}~\bibnamefont {Clarke}}, \bibinfo {author} {\bibfnamefont
  {M.}~\bibnamefont {M\"uck}}, \ and\ \bibinfo {author} {\bibfnamefont {R.~F.}\
  \bibnamefont {Bradley}},\ }\href {\doibase 10.1103/PhysRevD.64.092003}
  {\bibfield  {journal} {\bibinfo  {journal} {Phys. Rev. D}\ }\textbf {\bibinfo
  {volume} {64}},\ \bibinfo {pages} {092003} (\bibinfo {year}
  {2001})}\BibitemShut {NoStop}%
\bibitem [{\citenamefont {Asztalos}\ \emph
  {et~al.}(2010{\natexlab{b}})\citenamefont {Asztalos}, \citenamefont {Carosi},
  \citenamefont {Hagmann}, \citenamefont {Kinion}, \citenamefont {van Bibber},
  \citenamefont {Hotz}, \citenamefont {Rosenberg}, \citenamefont {Rybka},
  \citenamefont {Hoskins}, \citenamefont {Hwang}, \citenamefont {Sikivie},
  \citenamefont {Tanner}, \citenamefont {Bradley},\ and\ \citenamefont
  {Clarke}}]{asztalos_2010}%
  \BibitemOpen
  \bibfield  {author} {\bibinfo {author} {\bibfnamefont {S.~J.}\ \bibnamefont
  {Asztalos}}, \bibinfo {author} {\bibfnamefont {G.}~\bibnamefont {Carosi}},
  \bibinfo {author} {\bibfnamefont {C.}~\bibnamefont {Hagmann}}, \bibinfo
  {author} {\bibfnamefont {D.}~\bibnamefont {Kinion}}, \bibinfo {author}
  {\bibfnamefont {K.}~\bibnamefont {van Bibber}}, \bibinfo {author}
  {\bibfnamefont {M.}~\bibnamefont {Hotz}}, \bibinfo {author} {\bibfnamefont
  {L.~J.}\ \bibnamefont {Rosenberg}}, \bibinfo {author} {\bibfnamefont
  {G.}~\bibnamefont {Rybka}}, \bibinfo {author} {\bibfnamefont
  {J.}~\bibnamefont {Hoskins}}, \bibinfo {author} {\bibfnamefont
  {J.}~\bibnamefont {Hwang}}, \bibinfo {author} {\bibfnamefont
  {P.}~\bibnamefont {Sikivie}}, \bibinfo {author} {\bibfnamefont {D.~B.}\
  \bibnamefont {Tanner}}, \bibinfo {author} {\bibfnamefont {R.}~\bibnamefont
  {Bradley}}, \ and\ \bibinfo {author} {\bibfnamefont {J.}~\bibnamefont
  {Clarke}},\ }\href {\doibase 10.1103/PhysRevLett.104.041301} {\bibfield
  {journal} {\bibinfo  {journal} {Phys. Rev. Lett.}\ }\textbf {\bibinfo
  {volume} {104}},\ \bibinfo {pages} {041301} (\bibinfo {year}
  {2010}{\natexlab{b}})}\BibitemShut {NoStop}%
\bibitem [{\citenamefont {Du}\ \emph {et~al.}(2018)\citenamefont {Du},
  \citenamefont {Force}, \citenamefont {Khatiwada}, \citenamefont {Lentz},
  \citenamefont {Ottens}, \citenamefont {Rosenberg}, \citenamefont {Rybka},
  \citenamefont {Carosi}, \citenamefont {Woollett}, \citenamefont {Bowring},
  \citenamefont {Chou}, \citenamefont {Sonnenschein}, \citenamefont {Wester},
  \citenamefont {Boutan}, \citenamefont {Oblath}, \citenamefont {Bradley},
  \citenamefont {Daw}, \citenamefont {Dixit}, \citenamefont {Clarke},
  \citenamefont {O'Kelley}, \citenamefont {Crisosto}, \citenamefont {Gleason},
  \citenamefont {Jois}, \citenamefont {Sikivie}, \citenamefont {Stern},
  \citenamefont {Sullivan}, \citenamefont {Tanner},\ and\ \citenamefont
  {Hilton}}]{admx_2018}%
  \BibitemOpen
  \bibfield  {author} {\bibinfo {author} {\bibfnamefont {N.}~\bibnamefont
  {Du}}, \bibinfo {author} {\bibfnamefont {N.}~\bibnamefont {Force}}, \bibinfo
  {author} {\bibfnamefont {R.}~\bibnamefont {Khatiwada}}, \bibinfo {author}
  {\bibfnamefont {E.}~\bibnamefont {Lentz}}, \bibinfo {author} {\bibfnamefont
  {R.}~\bibnamefont {Ottens}}, \bibinfo {author} {\bibfnamefont {L.~J.}\
  \bibnamefont {Rosenberg}}, \bibinfo {author} {\bibfnamefont {G.}~\bibnamefont
  {Rybka}}, \bibinfo {author} {\bibfnamefont {G.}~\bibnamefont {Carosi}},
  \bibinfo {author} {\bibfnamefont {N.}~\bibnamefont {Woollett}}, \bibinfo
  {author} {\bibfnamefont {D.}~\bibnamefont {Bowring}}, \bibinfo {author}
  {\bibfnamefont {A.~S.}\ \bibnamefont {Chou}}, \bibinfo {author}
  {\bibfnamefont {A.}~\bibnamefont {Sonnenschein}}, \bibinfo {author}
  {\bibfnamefont {W.}~\bibnamefont {Wester}}, \bibinfo {author} {\bibfnamefont
  {C.}~\bibnamefont {Boutan}}, \bibinfo {author} {\bibfnamefont {N.~S.}\
  \bibnamefont {Oblath}}, \bibinfo {author} {\bibfnamefont {R.}~\bibnamefont
  {Bradley}}, \bibinfo {author} {\bibfnamefont {E.~J.}\ \bibnamefont {Daw}},
  \bibinfo {author} {\bibfnamefont {A.~V.}\ \bibnamefont {Dixit}}, \bibinfo
  {author} {\bibfnamefont {J.}~\bibnamefont {Clarke}}, \bibinfo {author}
  {\bibfnamefont {S.~R.}\ \bibnamefont {O'Kelley}}, \bibinfo {author}
  {\bibfnamefont {N.}~\bibnamefont {Crisosto}}, \bibinfo {author}
  {\bibfnamefont {J.~R.}\ \bibnamefont {Gleason}}, \bibinfo {author}
  {\bibfnamefont {S.}~\bibnamefont {Jois}}, \bibinfo {author} {\bibfnamefont
  {P.}~\bibnamefont {Sikivie}}, \bibinfo {author} {\bibfnamefont
  {I.}~\bibnamefont {Stern}}, \bibinfo {author} {\bibfnamefont {N.~S.}\
  \bibnamefont {Sullivan}}, \bibinfo {author} {\bibfnamefont {D.~B.}\
  \bibnamefont {Tanner}}, \ and\ \bibinfo {author} {\bibfnamefont {G.~C.}\
  \bibnamefont {Hilton}} (\bibinfo {collaboration} {ADMX Collaboration}),\
  }\href {\doibase 10.1103/PhysRevLett.120.151301} {\bibfield  {journal}
  {\bibinfo  {journal} {Phys. Rev. Lett.}\ }\textbf {\bibinfo {volume} {120}},\
  \bibinfo {pages} {151301} (\bibinfo {year} {2018})}\BibitemShut {NoStop}%
\bibitem [{\citenamefont {Vogel}\ \emph {et~al.}(2015)\citenamefont {Vogel},
  \citenamefont {Armengaud}, \citenamefont {Avignone}, \citenamefont {Betz},
  \citenamefont {Brax}, \citenamefont {Brun}, \citenamefont {Cantatore},
  \citenamefont {Carmona}, \citenamefont {Carosi}, \citenamefont {Caspers},
  \citenamefont {Caspi}, \citenamefont {Cetin}, \citenamefont {Chelouche},
  \citenamefont {Christensen}, \citenamefont {Dael}, \citenamefont {Dafni},
  \citenamefont {Davenport}, \citenamefont {Derbin}, \citenamefont {Desch},
  \citenamefont {Diago}, \citenamefont {D{\"o}brich}, \citenamefont
  {Dratchnev}, \citenamefont {Dudarev}, \citenamefont {Eleftheriadis},
  \citenamefont {Fanourakis}, \citenamefont {Ferrer-Ribas}, \citenamefont
  {Gal{\'a}n}, \citenamefont {Garc{\'i}a}, \citenamefont {Garza}, \citenamefont
  {Geralis}, \citenamefont {Gimeno}, \citenamefont {Giomataris}, \citenamefont
  {Gninenko}, \citenamefont {G{\'o}mez}, \citenamefont {Gonz{\'a}lez-D{\'i}az},
  \citenamefont {Guendelman}, \citenamefont {Hailey}, \citenamefont
  {Hiramatsu}, \citenamefont {Hoffmann}, \citenamefont {Horns}, \citenamefont
  {Iguaz}, \citenamefont {Irastorza}, \citenamefont {Isern}, \citenamefont
  {Imai}, \citenamefont {Jakobsen}, \citenamefont {Jaeckel}, \citenamefont
  {Jakovcic}, \citenamefont {Kaminski}, \citenamefont {Kawasaki}, \citenamefont
  {Karuza}, \citenamefont {Krcmar}, \citenamefont {Kousouris}, \citenamefont
  {Krieger}, \citenamefont {Lakic}, \citenamefont {Limousin}, \citenamefont
  {Lindner}, \citenamefont {Liolios}, \citenamefont {Luz{\'o}n}, \citenamefont
  {Matsuki}, \citenamefont {Muratova}, \citenamefont {Nones}, \citenamefont
  {Ortega}, \citenamefont {Papaevangelou}, \citenamefont {Pivovaroff},
  \citenamefont {Raffelt}, \citenamefont {Redondo}, \citenamefont {Ringwald},
  \citenamefont {Russenschuck}, \citenamefont {Ruz}, \citenamefont {Saikawa},
  \citenamefont {Savvidis}, \citenamefont {Sekiguchi}, \citenamefont
  {Semertzidis}, \citenamefont {Shilon}, \citenamefont {Sikivie}, \citenamefont
  {Silva}, \citenamefont {ten Kate}, \citenamefont {Tomas}, \citenamefont
  {Troitsky}, \citenamefont {Vafeiadis}, \citenamefont {van Bibber},
  \citenamefont {Vedrine}, \citenamefont {Villar}, \citenamefont {Walckiers},
  \citenamefont {Weltman}, \citenamefont {Wester}, \citenamefont {Yildiz},\
  and\ \citenamefont {Zioutas}}]{vogel_2015}%
  \BibitemOpen
  \bibfield  {author} {\bibinfo {author} {\bibfnamefont {J.~K.}\ \bibnamefont
  {Vogel}}, \bibinfo {author} {\bibfnamefont {E.}~\bibnamefont {Armengaud}},
  \bibinfo {author} {\bibfnamefont {F.~T.}\ \bibnamefont {Avignone}}, \bibinfo
  {author} {\bibfnamefont {M.}~\bibnamefont {Betz}}, \bibinfo {author}
  {\bibfnamefont {P.}~\bibnamefont {Brax}}, \bibinfo {author} {\bibfnamefont
  {P.}~\bibnamefont {Brun}}, \bibinfo {author} {\bibfnamefont {G.}~\bibnamefont
  {Cantatore}}, \bibinfo {author} {\bibfnamefont {J.~M.}\ \bibnamefont
  {Carmona}}, \bibinfo {author} {\bibfnamefont {G.~P.}\ \bibnamefont {Carosi}},
  \bibinfo {author} {\bibfnamefont {F.}~\bibnamefont {Caspers}}, \bibinfo
  {author} {\bibfnamefont {S.}~\bibnamefont {Caspi}}, \bibinfo {author}
  {\bibfnamefont {S.~A.}\ \bibnamefont {Cetin}}, \bibinfo {author}
  {\bibfnamefont {D.}~\bibnamefont {Chelouche}}, \bibinfo {author}
  {\bibfnamefont {F.~E.}\ \bibnamefont {Christensen}}, \bibinfo {author}
  {\bibfnamefont {A.}~\bibnamefont {Dael}}, \bibinfo {author} {\bibfnamefont
  {T.}~\bibnamefont {Dafni}}, \bibinfo {author} {\bibfnamefont
  {M.}~\bibnamefont {Davenport}}, \bibinfo {author} {\bibfnamefont {A.~V.}\
  \bibnamefont {Derbin}}, \bibinfo {author} {\bibfnamefont {K.}~\bibnamefont
  {Desch}}, \bibinfo {author} {\bibfnamefont {A.}~\bibnamefont {Diago}},
  \bibinfo {author} {\bibfnamefont {B.}~\bibnamefont {D{\"o}brich}}, \bibinfo
  {author} {\bibfnamefont {I.}~\bibnamefont {Dratchnev}}, \bibinfo {author}
  {\bibfnamefont {A.}~\bibnamefont {Dudarev}}, \bibinfo {author} {\bibfnamefont
  {C.}~\bibnamefont {Eleftheriadis}}, \bibinfo {author} {\bibfnamefont
  {G.}~\bibnamefont {Fanourakis}}, \bibinfo {author} {\bibfnamefont
  {E.}~\bibnamefont {Ferrer-Ribas}}, \bibinfo {author} {\bibfnamefont
  {J.}~\bibnamefont {Gal{\'a}n}}, \bibinfo {author} {\bibfnamefont {J.~A.}\
  \bibnamefont {Garc{\'i}a}}, \bibinfo {author} {\bibfnamefont {J.~G.}\
  \bibnamefont {Garza}}, \bibinfo {author} {\bibfnamefont {T.}~\bibnamefont
  {Geralis}}, \bibinfo {author} {\bibfnamefont {B.}~\bibnamefont {Gimeno}},
  \bibinfo {author} {\bibfnamefont {I.}~\bibnamefont {Giomataris}}, \bibinfo
  {author} {\bibfnamefont {S.}~\bibnamefont {Gninenko}}, \bibinfo {author}
  {\bibfnamefont {H.}~\bibnamefont {G{\'o}mez}}, \bibinfo {author}
  {\bibfnamefont {D.}~\bibnamefont {Gonz{\'a}lez-D{\'i}az}}, \bibinfo {author}
  {\bibfnamefont {E.}~\bibnamefont {Guendelman}}, \bibinfo {author}
  {\bibfnamefont {C.~J.}\ \bibnamefont {Hailey}}, \bibinfo {author}
  {\bibfnamefont {T.}~\bibnamefont {Hiramatsu}}, \bibinfo {author}
  {\bibfnamefont {D.~H.~H.}\ \bibnamefont {Hoffmann}}, \bibinfo {author}
  {\bibfnamefont {D.}~\bibnamefont {Horns}}, \bibinfo {author} {\bibfnamefont
  {F.~J.}\ \bibnamefont {Iguaz}}, \bibinfo {author} {\bibfnamefont {I.~G.}\
  \bibnamefont {Irastorza}}, \bibinfo {author} {\bibfnamefont {J.}~\bibnamefont
  {Isern}}, \bibinfo {author} {\bibfnamefont {K.}~\bibnamefont {Imai}},
  \bibinfo {author} {\bibfnamefont {A.~C.}\ \bibnamefont {Jakobsen}}, \bibinfo
  {author} {\bibfnamefont {J.}~\bibnamefont {Jaeckel}}, \bibinfo {author}
  {\bibfnamefont {K.}~\bibnamefont {Jakovcic}}, \bibinfo {author}
  {\bibfnamefont {J.}~\bibnamefont {Kaminski}}, \bibinfo {author}
  {\bibfnamefont {M.}~\bibnamefont {Kawasaki}}, \bibinfo {author}
  {\bibfnamefont {M.}~\bibnamefont {Karuza}}, \bibinfo {author} {\bibfnamefont
  {M.}~\bibnamefont {Krcmar}}, \bibinfo {author} {\bibfnamefont
  {K.}~\bibnamefont {Kousouris}}, \bibinfo {author} {\bibfnamefont
  {C.}~\bibnamefont {Krieger}}, \bibinfo {author} {\bibfnamefont
  {B.}~\bibnamefont {Lakic}}, \bibinfo {author} {\bibfnamefont
  {O.}~\bibnamefont {Limousin}}, \bibinfo {author} {\bibfnamefont
  {A.}~\bibnamefont {Lindner}}, \bibinfo {author} {\bibfnamefont
  {A.}~\bibnamefont {Liolios}}, \bibinfo {author} {\bibfnamefont
  {G.}~\bibnamefont {Luz{\'o}n}}, \bibinfo {author} {\bibfnamefont
  {S.}~\bibnamefont {Matsuki}}, \bibinfo {author} {\bibfnamefont {V.~N.}\
  \bibnamefont {Muratova}}, \bibinfo {author} {\bibfnamefont {C.}~\bibnamefont
  {Nones}}, \bibinfo {author} {\bibfnamefont {I.}~\bibnamefont {Ortega}},
  \bibinfo {author} {\bibfnamefont {T.}~\bibnamefont {Papaevangelou}}, \bibinfo
  {author} {\bibfnamefont {M.~J.}\ \bibnamefont {Pivovaroff}}, \bibinfo
  {author} {\bibfnamefont {G.}~\bibnamefont {Raffelt}}, \bibinfo {author}
  {\bibfnamefont {J.}~\bibnamefont {Redondo}}, \bibinfo {author} {\bibfnamefont
  {A.}~\bibnamefont {Ringwald}}, \bibinfo {author} {\bibfnamefont
  {S.}~\bibnamefont {Russenschuck}}, \bibinfo {author} {\bibfnamefont
  {J.}~\bibnamefont {Ruz}}, \bibinfo {author} {\bibfnamefont {K.}~\bibnamefont
  {Saikawa}}, \bibinfo {author} {\bibfnamefont {I.}~\bibnamefont {Savvidis}},
  \bibinfo {author} {\bibfnamefont {T.}~\bibnamefont {Sekiguchi}}, \bibinfo
  {author} {\bibfnamefont {Y.~K.}\ \bibnamefont {Semertzidis}}, \bibinfo
  {author} {\bibfnamefont {I.}~\bibnamefont {Shilon}}, \bibinfo {author}
  {\bibfnamefont {P.}~\bibnamefont {Sikivie}}, \bibinfo {author} {\bibfnamefont
  {H.}~\bibnamefont {Silva}}, \bibinfo {author} {\bibfnamefont
  {H.}~\bibnamefont {ten Kate}}, \bibinfo {author} {\bibfnamefont
  {A.}~\bibnamefont {Tomas}}, \bibinfo {author} {\bibfnamefont
  {S.}~\bibnamefont {Troitsky}}, \bibinfo {author} {\bibfnamefont
  {T.}~\bibnamefont {Vafeiadis}}, \bibinfo {author} {\bibfnamefont
  {K.}~\bibnamefont {van Bibber}}, \bibinfo {author} {\bibfnamefont
  {P.}~\bibnamefont {Vedrine}}, \bibinfo {author} {\bibfnamefont {J.~A.}\
  \bibnamefont {Villar}}, \bibinfo {author} {\bibfnamefont {L.}~\bibnamefont
  {Walckiers}}, \bibinfo {author} {\bibfnamefont {A.}~\bibnamefont {Weltman}},
  \bibinfo {author} {\bibfnamefont {W.}~\bibnamefont {Wester}}, \bibinfo
  {author} {\bibfnamefont {S.~C.}\ \bibnamefont {Yildiz}}, \ and\ \bibinfo
  {author} {\bibfnamefont {K.}~\bibnamefont {Zioutas}},\ }\href
  {http://www.sciencedirect.com/science/article/pii/S1875389214006440}
  {\bibfield  {journal} {\bibinfo  {journal} {Physics Procedia}\ }\textbf
  {\bibinfo {volume} {61}},\ \bibinfo {pages} {193} (\bibinfo {year}
  {2015})}\BibitemShut {NoStop}%
\bibitem [{\citenamefont {Caldwell}\ \emph {et~al.}(2017)\citenamefont
  {Caldwell}, \citenamefont {Dvali}, \citenamefont {Majorovits}, \citenamefont
  {Millar}, \citenamefont {Raffelt}, \citenamefont {Redondo}, \citenamefont
  {Reimann}, \citenamefont {Simon},\ and\ \citenamefont
  {Steffen}}]{caldwell_2017}%
  \BibitemOpen
  \bibfield  {author} {\bibinfo {author} {\bibfnamefont {A.}~\bibnamefont
  {Caldwell}}, \bibinfo {author} {\bibfnamefont {G.}~\bibnamefont {Dvali}},
  \bibinfo {author} {\bibfnamefont {B.}~\bibnamefont {Majorovits}}, \bibinfo
  {author} {\bibfnamefont {A.}~\bibnamefont {Millar}}, \bibinfo {author}
  {\bibfnamefont {G.}~\bibnamefont {Raffelt}}, \bibinfo {author} {\bibfnamefont
  {J.}~\bibnamefont {Redondo}}, \bibinfo {author} {\bibfnamefont
  {O.}~\bibnamefont {Reimann}}, \bibinfo {author} {\bibfnamefont
  {F.}~\bibnamefont {Simon}}, \ and\ \bibinfo {author} {\bibfnamefont
  {F.}~\bibnamefont {Steffen}} (\bibinfo {collaboration} {MADMAX Working
  Group}),\ }\href {\doibase 10.1103/PhysRevLett.118.091801} {\bibfield
  {journal} {\bibinfo  {journal} {Phys. Rev. Lett.}\ }\textbf {\bibinfo
  {volume} {118}},\ \bibinfo {pages} {091801} (\bibinfo {year}
  {2017})}\BibitemShut {NoStop}%
\bibitem [{\citenamefont {Januschek}(2014)}]{friederike_2014}%
  \BibitemOpen
  \bibfield  {author} {\bibinfo {author} {\bibfnamefont {F.}~\bibnamefont
  {Januschek}},\ }\href {\doibase
  10.3204/desy-proc-2014-03/januschek_friederike} {\bibfield  {journal}
  {\bibinfo  {journal} {Proceedings of the 10th Patras Workshop on Axions}\
  }\textbf {\bibinfo {volume} {1}},\ \bibinfo {pages} {83} (\bibinfo {year}
  {2014})}\BibitemShut {NoStop}%
\bibitem [{\citenamefont {Bähre}\ \emph {et~al.}(2013)\citenamefont {Bähre},
  \citenamefont {Döbrich}, \citenamefont {Dreyling-Eschweiler}, \citenamefont
  {Ghazaryan}, \citenamefont {Hodajerdi}, \citenamefont {Horns}, \citenamefont
  {Januschek}, \citenamefont {Knabbe}, \citenamefont {Lindner}, \citenamefont
  {Notz}, \citenamefont {Ringwald}, \citenamefont {von Seggern}, \citenamefont
  {Stromhagen}, \citenamefont {Trines},\ and\ \citenamefont
  {Willke}}]{baehre_2013}%
  \BibitemOpen
  \bibfield  {author} {\bibinfo {author} {\bibfnamefont {R.}~\bibnamefont
  {Bähre}}, \bibinfo {author} {\bibfnamefont {B.}~\bibnamefont {Döbrich}},
  \bibinfo {author} {\bibfnamefont {J.}~\bibnamefont {Dreyling-Eschweiler}},
  \bibinfo {author} {\bibfnamefont {S.}~\bibnamefont {Ghazaryan}}, \bibinfo
  {author} {\bibfnamefont {R.}~\bibnamefont {Hodajerdi}}, \bibinfo {author}
  {\bibfnamefont {D.}~\bibnamefont {Horns}}, \bibinfo {author} {\bibfnamefont
  {F.}~\bibnamefont {Januschek}}, \bibinfo {author} {\bibfnamefont {E.~A.}\
  \bibnamefont {Knabbe}}, \bibinfo {author} {\bibfnamefont {A.}~\bibnamefont
  {Lindner}}, \bibinfo {author} {\bibfnamefont {D.}~\bibnamefont {Notz}},
  \bibinfo {author} {\bibfnamefont {A.}~\bibnamefont {Ringwald}}, \bibinfo
  {author} {\bibfnamefont {J.~E.}\ \bibnamefont {von Seggern}}, \bibinfo
  {author} {\bibfnamefont {R.}~\bibnamefont {Stromhagen}}, \bibinfo {author}
  {\bibfnamefont {D.}~\bibnamefont {Trines}}, \ and\ \bibinfo {author}
  {\bibfnamefont {B.}~\bibnamefont {Willke}},\ }\href {\doibase
  10.1088/1748-0221/8/09/t09001} {\bibfield  {journal} {\bibinfo  {journal}
  {Journal of Instrumentation}\ }\textbf {\bibinfo {volume} {8}},\ \bibinfo
  {pages} {T09001} (\bibinfo {year} {2013})}\BibitemShut {NoStop}%
\bibitem [{\citenamefont {Pugnat}\ \emph {et~al.}(2008)\citenamefont {Pugnat},
  \citenamefont {Duvillaret}, \citenamefont {Jost}, \citenamefont {Vitrant},
  \citenamefont {Romanini}, \citenamefont {Siemko}, \citenamefont {Ballou},
  \citenamefont {Barbara}, \citenamefont {Finger}, \citenamefont {Finger},
  \citenamefont {Ho\ifmmode~\check{s}\else \v{s}\fi{}ek}, \citenamefont
  {Kr\'al}, \citenamefont {Meissner}, \citenamefont {\ifmmode~\check{S}\else
  \v{S}\fi{}ulc},\ and\ \citenamefont {Zicha}}]{osqar_2008}%
  \BibitemOpen
  \bibfield  {author} {\bibinfo {author} {\bibfnamefont {P.}~\bibnamefont
  {Pugnat}}, \bibinfo {author} {\bibfnamefont {L.}~\bibnamefont {Duvillaret}},
  \bibinfo {author} {\bibfnamefont {R.}~\bibnamefont {Jost}}, \bibinfo {author}
  {\bibfnamefont {G.}~\bibnamefont {Vitrant}}, \bibinfo {author} {\bibfnamefont
  {D.}~\bibnamefont {Romanini}}, \bibinfo {author} {\bibfnamefont
  {A.}~\bibnamefont {Siemko}}, \bibinfo {author} {\bibfnamefont
  {R.}~\bibnamefont {Ballou}}, \bibinfo {author} {\bibfnamefont
  {B.}~\bibnamefont {Barbara}}, \bibinfo {author} {\bibfnamefont
  {M.}~\bibnamefont {Finger}}, \bibinfo {author} {\bibfnamefont
  {M.}~\bibnamefont {Finger}}, \bibinfo {author} {\bibfnamefont
  {J.}~\bibnamefont {Ho\ifmmode~\check{s}\else \v{s}\fi{}ek}}, \bibinfo
  {author} {\bibfnamefont {M.}~\bibnamefont {Kr\'al}}, \bibinfo {author}
  {\bibfnamefont {K.~A.}\ \bibnamefont {Meissner}}, \bibinfo {author}
  {\bibfnamefont {M.}~\bibnamefont {\ifmmode~\check{S}\else \v{S}\fi{}ulc}}, \
  and\ \bibinfo {author} {\bibfnamefont {J.}~\bibnamefont {Zicha}} (\bibinfo
  {collaboration} {OSQAR Collaboration}),\ }\href {\doibase
  10.1103/PhysRevD.78.092003} {\bibfield  {journal} {\bibinfo  {journal} {Phys.
  Rev. D}\ }\textbf {\bibinfo {volume} {78}},\ \bibinfo {pages} {092003}
  (\bibinfo {year} {2008})}\BibitemShut {NoStop}%
\bibitem [{\citenamefont {Capparelli}\ \emph {et~al.}(2016)\citenamefont
  {Capparelli}, \citenamefont {Cavoto}, \citenamefont {Ferretti}, \citenamefont
  {Giazotto}, \citenamefont {Polosa},\ and\ \citenamefont
  {Spagnolo}}]{capparelli_2016}%
  \BibitemOpen
  \bibfield  {author} {\bibinfo {author} {\bibfnamefont {L.}~\bibnamefont
  {Capparelli}}, \bibinfo {author} {\bibfnamefont {G.}~\bibnamefont {Cavoto}},
  \bibinfo {author} {\bibfnamefont {J.}~\bibnamefont {Ferretti}}, \bibinfo
  {author} {\bibfnamefont {F.}~\bibnamefont {Giazotto}}, \bibinfo {author}
  {\bibfnamefont {A.}~\bibnamefont {Polosa}}, \ and\ \bibinfo {author}
  {\bibfnamefont {P.}~\bibnamefont {Spagnolo}},\ }\href {\doibase
  https://doi.org/10.1016/j.dark.2016.01.003} {\bibfield  {journal} {\bibinfo
  {journal} {Physics of the Dark Universe}\ }\textbf {\bibinfo {volume} {12}},\
  \bibinfo {pages} {37 } (\bibinfo {year} {2016})}\BibitemShut {NoStop}%
\bibitem [{\citenamefont {Betz}\ \emph {et~al.}(2013)\citenamefont {Betz},
  \citenamefont {Caspers}, \citenamefont {Gasior}, \citenamefont {Thumm},\ and\
  \citenamefont {Rieger}}]{betz_2013}%
  \BibitemOpen
  \bibfield  {author} {\bibinfo {author} {\bibfnamefont {M.}~\bibnamefont
  {Betz}}, \bibinfo {author} {\bibfnamefont {F.}~\bibnamefont {Caspers}},
  \bibinfo {author} {\bibfnamefont {M.}~\bibnamefont {Gasior}}, \bibinfo
  {author} {\bibfnamefont {M.}~\bibnamefont {Thumm}}, \ and\ \bibinfo {author}
  {\bibfnamefont {S.~W.}\ \bibnamefont {Rieger}},\ }\href {\doibase
  10.1103/PhysRevD.88.075014} {\bibfield  {journal} {\bibinfo  {journal} {Phys.
  Rev. D}\ }\textbf {\bibinfo {volume} {88}},\ \bibinfo {pages} {075014}
  (\bibinfo {year} {2013})}\BibitemShut {NoStop}%
\bibitem [{\citenamefont {Raffelt}\ and\ \citenamefont
  {Stodolsky}(1988)}]{raffelt_1988}%
  \BibitemOpen
  \bibfield  {author} {\bibinfo {author} {\bibfnamefont {G.}~\bibnamefont
  {Raffelt}}\ and\ \bibinfo {author} {\bibfnamefont {L.}~\bibnamefont
  {Stodolsky}},\ }\href {\doibase 10.1103/PhysRevD.37.1237} {\bibfield
  {journal} {\bibinfo  {journal} {Phys. Rev. D}\ }\textbf {\bibinfo {volume}
  {37}},\ \bibinfo {pages} {1237} (\bibinfo {year} {1988})}\BibitemShut
  {NoStop}%
\bibitem [{\citenamefont {Mangles}\ \emph {et~al.}(2004)\citenamefont
  {Mangles}, \citenamefont {Murphy}, \citenamefont {Najmudin}, \citenamefont
  {Thomas}, \citenamefont {Collier}, \citenamefont {Dangor}, \citenamefont
  {Divall}, \citenamefont {Foster}, \citenamefont {Gallacher}, \citenamefont
  {Hooker}, \citenamefont {Jaroszynski}, \citenamefont {Langley}, \citenamefont
  {Mori}, \citenamefont {Norreys}, \citenamefont {Tsung}, \citenamefont
  {Viskup}, \citenamefont {Walton},\ and\ \citenamefont
  {Krushelnick}}]{ee183cdbe0634a8dbaabf276f03a1d91}%
  \BibitemOpen
  \bibfield  {author} {\bibinfo {author} {\bibfnamefont {S.}~\bibnamefont
  {Mangles}}, \bibinfo {author} {\bibfnamefont {C.}~\bibnamefont {Murphy}},
  \bibinfo {author} {\bibfnamefont {Z.}~\bibnamefont {Najmudin}}, \bibinfo
  {author} {\bibfnamefont {A.}~\bibnamefont {Thomas}}, \bibinfo {author}
  {\bibfnamefont {J.}~\bibnamefont {Collier}}, \bibinfo {author} {\bibfnamefont
  {A.}~\bibnamefont {Dangor}}, \bibinfo {author} {\bibfnamefont
  {E.}~\bibnamefont {Divall}}, \bibinfo {author} {\bibfnamefont
  {P.}~\bibnamefont {Foster}}, \bibinfo {author} {\bibfnamefont
  {J.}~\bibnamefont {Gallacher}}, \bibinfo {author} {\bibfnamefont
  {C.}~\bibnamefont {Hooker}}, \bibinfo {author} {\bibfnamefont
  {D.}~\bibnamefont {Jaroszynski}}, \bibinfo {author} {\bibfnamefont
  {A.}~\bibnamefont {Langley}}, \bibinfo {author} {\bibfnamefont
  {W.}~\bibnamefont {Mori}}, \bibinfo {author} {\bibfnamefont {P.}~\bibnamefont
  {Norreys}}, \bibinfo {author} {\bibfnamefont {F.}~\bibnamefont {Tsung}},
  \bibinfo {author} {\bibfnamefont {R.}~\bibnamefont {Viskup}}, \bibinfo
  {author} {\bibfnamefont {B.}~\bibnamefont {Walton}}, \ and\ \bibinfo {author}
  {\bibfnamefont {K.}~\bibnamefont {Krushelnick}},\ }\href {\doibase
  10.1038/nature02939} {\bibfield  {journal} {\bibinfo  {journal} {Nature}\
  }\textbf {\bibinfo {volume} {431}},\ \bibinfo {pages} {535} (\bibinfo {year}
  {2004})}\BibitemShut {NoStop}%
\bibitem [{\citenamefont {Geddes}\ \emph {et~al.}(2004)\citenamefont {Geddes},
  \citenamefont {Toth}, \citenamefont {van Tilborg}, \citenamefont {Esarey},
  \citenamefont {Schroeder}, \citenamefont {Bruhwiler}, \citenamefont {Nieter},
  \citenamefont {Cary},\ and\ \citenamefont {Leemans}}]{Geddes2004}%
  \BibitemOpen
  \bibfield  {author} {\bibinfo {author} {\bibfnamefont {C.~G.~R.}\
  \bibnamefont {Geddes}}, \bibinfo {author} {\bibfnamefont {C.}~\bibnamefont
  {Toth}}, \bibinfo {author} {\bibfnamefont {J.}~\bibnamefont {van Tilborg}},
  \bibinfo {author} {\bibfnamefont {E.}~\bibnamefont {Esarey}}, \bibinfo
  {author} {\bibfnamefont {C.~B.}\ \bibnamefont {Schroeder}}, \bibinfo {author}
  {\bibfnamefont {D.}~\bibnamefont {Bruhwiler}}, \bibinfo {author}
  {\bibfnamefont {C.}~\bibnamefont {Nieter}}, \bibinfo {author} {\bibfnamefont
  {J.}~\bibnamefont {Cary}}, \ and\ \bibinfo {author} {\bibfnamefont {W.~P.}\
  \bibnamefont {Leemans}},\ }\href {http://dx.doi.org/10.1038/nature02900}
  {\bibfield  {journal} {\bibinfo  {journal} {Nature}\ }\textbf {\bibinfo
  {volume} {431}},\ \bibinfo {pages} {538 EP } (\bibinfo {year}
  {2004})}\BibitemShut {NoStop}%
\bibitem [{\citenamefont {Faure}\ \emph {et~al.}(2004)\citenamefont {Faure},
  \citenamefont {Glinec}, \citenamefont {Pukhov}, \citenamefont {Kiselev},
  \citenamefont {Gordienko}, \citenamefont {Lefebvre}, \citenamefont
  {Rousseau}, \citenamefont {Burgy},\ and\ \citenamefont {Malka}}]{Faure2004}%
  \BibitemOpen
  \bibfield  {author} {\bibinfo {author} {\bibfnamefont {J.}~\bibnamefont
  {Faure}}, \bibinfo {author} {\bibfnamefont {Y.}~\bibnamefont {Glinec}},
  \bibinfo {author} {\bibfnamefont {A.}~\bibnamefont {Pukhov}}, \bibinfo
  {author} {\bibfnamefont {S.}~\bibnamefont {Kiselev}}, \bibinfo {author}
  {\bibfnamefont {S.}~\bibnamefont {Gordienko}}, \bibinfo {author}
  {\bibfnamefont {E.}~\bibnamefont {Lefebvre}}, \bibinfo {author}
  {\bibfnamefont {J.-P.}\ \bibnamefont {Rousseau}}, \bibinfo {author}
  {\bibfnamefont {F.}~\bibnamefont {Burgy}}, \ and\ \bibinfo {author}
  {\bibfnamefont {V.}~\bibnamefont {Malka}},\ }\href
  {http://dx.doi.org/10.1038/nature02963} {\bibfield  {journal} {\bibinfo
  {journal} {Nature}\ }\textbf {\bibinfo {volume} {431}},\ \bibinfo {pages}
  {541 EP } (\bibinfo {year} {2004})}\BibitemShut {NoStop}%
\bibitem [{\citenamefont {Adli}\ \emph {et~al.}(2018)\citenamefont {Adli},
  \citenamefont {Ahuja}, \citenamefont {Apsimon}, \citenamefont {Apsimon},
  \citenamefont {Bachmann}, \citenamefont {Barrientos}, \citenamefont {Batsch},
  \citenamefont {Bauche}, \citenamefont {Berglyd~Olsen}, \citenamefont
  {Bernardini}, \citenamefont {Bohl}, \citenamefont {Bracco}, \citenamefont
  {Braunm{\"u}ller}, \citenamefont {Burt}, \citenamefont {Buttensch{\"o}n},
  \citenamefont {Caldwell}, \citenamefont {Cascella}, \citenamefont {Chappell},
  \citenamefont {Chevallay}, \citenamefont {Chung}, \citenamefont {Cooke},
  \citenamefont {Damerau}, \citenamefont {Deacon}, \citenamefont {Deubner},
  \citenamefont {Dexter}, \citenamefont {Doebert}, \citenamefont {Farmer},
  \citenamefont {Fedosseev}, \citenamefont {Fiorito}, \citenamefont {Fonseca},
  \citenamefont {Friebel}, \citenamefont {Garolfi}, \citenamefont {Gessner},
  \citenamefont {Gorgisyan}, \citenamefont {Gorn}, \citenamefont {Granados},
  \citenamefont {Grulke}, \citenamefont {Gschwendtner}, \citenamefont {Hansen},
  \citenamefont {Helm}, \citenamefont {Henderson}, \citenamefont {H{\"u}ther},
  \citenamefont {Ibison}, \citenamefont {Jensen}, \citenamefont {Jolly},
  \citenamefont {Keeble}, \citenamefont {Kim}, \citenamefont {Kraus},
  \citenamefont {Li}, \citenamefont {Liu}, \citenamefont {Lopes}, \citenamefont
  {Lotov}, \citenamefont {Maricalva~Brun}, \citenamefont {Martyanov},
  \citenamefont {Mazzoni}, \citenamefont {Medina~Godoy}, \citenamefont
  {Minakov}, \citenamefont {Mitchell}, \citenamefont {Molendijk}, \citenamefont
  {Moody}, \citenamefont {Moreira}, \citenamefont {Muggli}, \citenamefont
  {{\"O}z}, \citenamefont {Pasquino}, \citenamefont {Pardons}, \citenamefont
  {Pe{\~n}a~Asmus}, \citenamefont {Pepitone}, \citenamefont {Perera},
  \citenamefont {Petrenko}, \citenamefont {Pitman}, \citenamefont {Pukhov},
  \citenamefont {Rey}, \citenamefont {Rieger}, \citenamefont {Ruhl},
  \citenamefont {Schmidt}, \citenamefont {Shalimova}, \citenamefont {Sherwood},
  \citenamefont {Silva}, \citenamefont {Soby}, \citenamefont {Sosedkin},
  \citenamefont {Speroni}, \citenamefont {Spitsyn}, \citenamefont {Tuev},
  \citenamefont {Turner}, \citenamefont {Velotti}, \citenamefont {Verra},
  \citenamefont {Verzilov}, \citenamefont {Vieira}, \citenamefont {Welsch},
  \citenamefont {Williamson}, \citenamefont {Wing}, \citenamefont {Woolley},\
  and\ \citenamefont {Xia}}]{adli_2018}%
  \BibitemOpen
  \bibfield  {author} {\bibinfo {author} {\bibfnamefont {E.}~\bibnamefont
  {Adli}}, \bibinfo {author} {\bibfnamefont {A.}~\bibnamefont {Ahuja}},
  \bibinfo {author} {\bibfnamefont {O.}~\bibnamefont {Apsimon}}, \bibinfo
  {author} {\bibfnamefont {R.}~\bibnamefont {Apsimon}}, \bibinfo {author}
  {\bibfnamefont {A.-M.}\ \bibnamefont {Bachmann}}, \bibinfo {author}
  {\bibfnamefont {D.}~\bibnamefont {Barrientos}}, \bibinfo {author}
  {\bibfnamefont {F.}~\bibnamefont {Batsch}}, \bibinfo {author} {\bibfnamefont
  {J.}~\bibnamefont {Bauche}}, \bibinfo {author} {\bibfnamefont {V.~K.}\
  \bibnamefont {Berglyd~Olsen}}, \bibinfo {author} {\bibfnamefont
  {M.}~\bibnamefont {Bernardini}}, \bibinfo {author} {\bibfnamefont
  {T.}~\bibnamefont {Bohl}}, \bibinfo {author} {\bibfnamefont {C.}~\bibnamefont
  {Bracco}}, \bibinfo {author} {\bibfnamefont {F.}~\bibnamefont
  {Braunm{\"u}ller}}, \bibinfo {author} {\bibfnamefont {G.}~\bibnamefont
  {Burt}}, \bibinfo {author} {\bibfnamefont {B.}~\bibnamefont
  {Buttensch{\"o}n}}, \bibinfo {author} {\bibfnamefont {A.}~\bibnamefont
  {Caldwell}}, \bibinfo {author} {\bibfnamefont {M.}~\bibnamefont {Cascella}},
  \bibinfo {author} {\bibfnamefont {J.}~\bibnamefont {Chappell}}, \bibinfo
  {author} {\bibfnamefont {E.}~\bibnamefont {Chevallay}}, \bibinfo {author}
  {\bibfnamefont {M.}~\bibnamefont {Chung}}, \bibinfo {author} {\bibfnamefont
  {D.}~\bibnamefont {Cooke}}, \bibinfo {author} {\bibfnamefont
  {H.}~\bibnamefont {Damerau}}, \bibinfo {author} {\bibfnamefont
  {L.}~\bibnamefont {Deacon}}, \bibinfo {author} {\bibfnamefont {L.~H.}\
  \bibnamefont {Deubner}}, \bibinfo {author} {\bibfnamefont {A.}~\bibnamefont
  {Dexter}}, \bibinfo {author} {\bibfnamefont {S.}~\bibnamefont {Doebert}},
  \bibinfo {author} {\bibfnamefont {J.}~\bibnamefont {Farmer}}, \bibinfo
  {author} {\bibfnamefont {V.~N.}\ \bibnamefont {Fedosseev}}, \bibinfo {author}
  {\bibfnamefont {R.}~\bibnamefont {Fiorito}}, \bibinfo {author} {\bibfnamefont
  {R.~A.}\ \bibnamefont {Fonseca}}, \bibinfo {author} {\bibfnamefont
  {F.}~\bibnamefont {Friebel}}, \bibinfo {author} {\bibfnamefont
  {L.}~\bibnamefont {Garolfi}}, \bibinfo {author} {\bibfnamefont
  {S.}~\bibnamefont {Gessner}}, \bibinfo {author} {\bibfnamefont
  {I.}~\bibnamefont {Gorgisyan}}, \bibinfo {author} {\bibfnamefont {A.~A.}\
  \bibnamefont {Gorn}}, \bibinfo {author} {\bibfnamefont {E.}~\bibnamefont
  {Granados}}, \bibinfo {author} {\bibfnamefont {O.}~\bibnamefont {Grulke}},
  \bibinfo {author} {\bibfnamefont {E.}~\bibnamefont {Gschwendtner}}, \bibinfo
  {author} {\bibfnamefont {J.}~\bibnamefont {Hansen}}, \bibinfo {author}
  {\bibfnamefont {A.}~\bibnamefont {Helm}}, \bibinfo {author} {\bibfnamefont
  {J.~R.}\ \bibnamefont {Henderson}}, \bibinfo {author} {\bibfnamefont
  {M.}~\bibnamefont {H{\"u}ther}}, \bibinfo {author} {\bibfnamefont
  {M.}~\bibnamefont {Ibison}}, \bibinfo {author} {\bibfnamefont
  {L.}~\bibnamefont {Jensen}}, \bibinfo {author} {\bibfnamefont
  {S.}~\bibnamefont {Jolly}}, \bibinfo {author} {\bibfnamefont
  {F.}~\bibnamefont {Keeble}}, \bibinfo {author} {\bibfnamefont {S.-Y.}\
  \bibnamefont {Kim}}, \bibinfo {author} {\bibfnamefont {F.}~\bibnamefont
  {Kraus}}, \bibinfo {author} {\bibfnamefont {Y.}~\bibnamefont {Li}}, \bibinfo
  {author} {\bibfnamefont {S.}~\bibnamefont {Liu}}, \bibinfo {author}
  {\bibfnamefont {N.}~\bibnamefont {Lopes}}, \bibinfo {author} {\bibfnamefont
  {K.~V.}\ \bibnamefont {Lotov}}, \bibinfo {author} {\bibfnamefont
  {L.}~\bibnamefont {Maricalva~Brun}}, \bibinfo {author} {\bibfnamefont
  {M.}~\bibnamefont {Martyanov}}, \bibinfo {author} {\bibfnamefont
  {S.}~\bibnamefont {Mazzoni}}, \bibinfo {author} {\bibfnamefont
  {D.}~\bibnamefont {Medina~Godoy}}, \bibinfo {author} {\bibfnamefont {V.~A.}\
  \bibnamefont {Minakov}}, \bibinfo {author} {\bibfnamefont {J.}~\bibnamefont
  {Mitchell}}, \bibinfo {author} {\bibfnamefont {J.~C.}\ \bibnamefont
  {Molendijk}}, \bibinfo {author} {\bibfnamefont {J.~T.}\ \bibnamefont
  {Moody}}, \bibinfo {author} {\bibfnamefont {M.}~\bibnamefont {Moreira}},
  \bibinfo {author} {\bibfnamefont {P.}~\bibnamefont {Muggli}}, \bibinfo
  {author} {\bibfnamefont {E.}~\bibnamefont {{\"O}z}}, \bibinfo {author}
  {\bibfnamefont {C.}~\bibnamefont {Pasquino}}, \bibinfo {author}
  {\bibfnamefont {A.}~\bibnamefont {Pardons}}, \bibinfo {author} {\bibfnamefont
  {F.}~\bibnamefont {Pe{\~n}a~Asmus}}, \bibinfo {author} {\bibfnamefont
  {K.}~\bibnamefont {Pepitone}}, \bibinfo {author} {\bibfnamefont
  {A.}~\bibnamefont {Perera}}, \bibinfo {author} {\bibfnamefont
  {A.}~\bibnamefont {Petrenko}}, \bibinfo {author} {\bibfnamefont
  {S.}~\bibnamefont {Pitman}}, \bibinfo {author} {\bibfnamefont
  {A.}~\bibnamefont {Pukhov}}, \bibinfo {author} {\bibfnamefont
  {S.}~\bibnamefont {Rey}}, \bibinfo {author} {\bibfnamefont {K.}~\bibnamefont
  {Rieger}}, \bibinfo {author} {\bibfnamefont {H.}~\bibnamefont {Ruhl}},
  \bibinfo {author} {\bibfnamefont {J.~S.}\ \bibnamefont {Schmidt}}, \bibinfo
  {author} {\bibfnamefont {I.~A.}\ \bibnamefont {Shalimova}}, \bibinfo {author}
  {\bibfnamefont {P.}~\bibnamefont {Sherwood}}, \bibinfo {author}
  {\bibfnamefont {L.~O.}\ \bibnamefont {Silva}}, \bibinfo {author}
  {\bibfnamefont {L.}~\bibnamefont {Soby}}, \bibinfo {author} {\bibfnamefont
  {A.~P.}\ \bibnamefont {Sosedkin}}, \bibinfo {author} {\bibfnamefont
  {R.}~\bibnamefont {Speroni}}, \bibinfo {author} {\bibfnamefont {R.~I.}\
  \bibnamefont {Spitsyn}}, \bibinfo {author} {\bibfnamefont {P.~V.}\
  \bibnamefont {Tuev}}, \bibinfo {author} {\bibfnamefont {M.}~\bibnamefont
  {Turner}}, \bibinfo {author} {\bibfnamefont {F.}~\bibnamefont {Velotti}},
  \bibinfo {author} {\bibfnamefont {L.}~\bibnamefont {Verra}}, \bibinfo
  {author} {\bibfnamefont {V.~A.}\ \bibnamefont {Verzilov}}, \bibinfo {author}
  {\bibfnamefont {J.}~\bibnamefont {Vieira}}, \bibinfo {author} {\bibfnamefont
  {C.~P.}\ \bibnamefont {Welsch}}, \bibinfo {author} {\bibfnamefont
  {B.}~\bibnamefont {Williamson}}, \bibinfo {author} {\bibfnamefont
  {M.}~\bibnamefont {Wing}}, \bibinfo {author} {\bibfnamefont {B.}~\bibnamefont
  {Woolley}}, \ and\ \bibinfo {author} {\bibfnamefont {G.}~\bibnamefont
  {Xia}},\ }\href {\doibase 10.1038/s41586-018-0485-4} {\bibfield  {journal}
  {\bibinfo  {journal} {Nature}\ }\textbf {\bibinfo {volume} {561}},\ \bibinfo
  {pages} {363} (\bibinfo {year} {2018})}\BibitemShut {NoStop}%
\bibitem [{\citenamefont {Burton}\ and\ \citenamefont
  {Noble}(2017)}]{1710.01906}%
  \BibitemOpen
  \bibfield  {author} {\bibinfo {author} {\bibfnamefont {D.~A.}\ \bibnamefont
  {Burton}}\ and\ \bibinfo {author} {\bibfnamefont {A.}~\bibnamefont {Noble}},\
  }\href@noop {} {\enquote {\bibinfo {title} {Plasma-based wakefield
  accelerators as sources of axion-like particles},}\ } (\bibinfo {year}
  {2017}),\ \Eprint {http://arxiv.org/abs/arXiv:1710.01906} {arXiv:1710.01906}
  \BibitemShut {NoStop}%
\bibitem [{\citenamefont {Burton}\ and\ \citenamefont
  {Noble}(2010)}]{1751-8121-43-7-075502}%
  \BibitemOpen
  \bibfield  {author} {\bibinfo {author} {\bibfnamefont {D.~A.}\ \bibnamefont
  {Burton}}\ and\ \bibinfo {author} {\bibfnamefont {A.}~\bibnamefont {Noble}},\
  }\href {http://stacks.iop.org/1751-8121/43/i=7/a=075502} {\bibfield
  {journal} {\bibinfo  {journal} {Journal of Physics A: Mathematical and
  Theoretical}\ }\textbf {\bibinfo {volume} {43}},\ \bibinfo {pages} {075502}
  (\bibinfo {year} {2010})}\BibitemShut {NoStop}%
\bibitem [{\citenamefont {Burton}\ \emph {et~al.}(2016)\citenamefont {Burton},
  \citenamefont {Noble},\ and\ \citenamefont
  {Walton}}]{1751-8121-49-38-385501}%
  \BibitemOpen
  \bibfield  {author} {\bibinfo {author} {\bibfnamefont {D.~A.}\ \bibnamefont
  {Burton}}, \bibinfo {author} {\bibfnamefont {A.}~\bibnamefont {Noble}}, \
  and\ \bibinfo {author} {\bibfnamefont {T.~J.}\ \bibnamefont {Walton}},\
  }\href {http://stacks.iop.org/1751-8121/49/i=38/a=385501} {\bibfield
  {journal} {\bibinfo  {journal} {Journal of Physics A: Mathematical and
  Theoretical}\ }\textbf {\bibinfo {volume} {49}},\ \bibinfo {pages} {385501}
  (\bibinfo {year} {2016})}\BibitemShut {NoStop}%
\bibitem [{\citenamefont {Mendon\c{c}a}(2007)}]{0295-5075-79-2-21001}%
  \BibitemOpen
  \bibfield  {author} {\bibinfo {author} {\bibfnamefont {J.~T.}\ \bibnamefont
  {Mendon\c{c}a}},\ }\href {http://stacks.iop.org/0295-5075/79/i=2/a=21001}
  {\bibfield  {journal} {\bibinfo  {journal} {EPL (Europhysics Letters)}\
  }\textbf {\bibinfo {volume} {79}},\ \bibinfo {pages} {21001} (\bibinfo {year}
  {2007})}\BibitemShut {NoStop}%
\bibitem [{\citenamefont {Pshirkov}\ and\ \citenamefont
  {Popov}(2009)}]{pshirkov_2009}%
  \BibitemOpen
  \bibfield  {author} {\bibinfo {author} {\bibfnamefont {M.~S.}\ \bibnamefont
  {Pshirkov}}\ and\ \bibinfo {author} {\bibfnamefont {S.~B.}\ \bibnamefont
  {Popov}},\ }\href {\doibase 10.1134/s1063776109030030} {\bibfield  {journal}
  {\bibinfo  {journal} {Journal of Experimental and Theoretical Physics}\
  }\textbf {\bibinfo {volume} {108}},\ \bibinfo {pages} {384} (\bibinfo {year}
  {2009})}\BibitemShut {NoStop}%
\bibitem [{\citenamefont {Hook}\ \emph {et~al.}(2018)\citenamefont {Hook},
  \citenamefont {Kahn}, \citenamefont {Safdi},\ and\ \citenamefont
  {Sun}}]{hook_2018}%
  \BibitemOpen
  \bibfield  {author} {\bibinfo {author} {\bibfnamefont {A.}~\bibnamefont
  {Hook}}, \bibinfo {author} {\bibfnamefont {Y.}~\bibnamefont {Kahn}}, \bibinfo
  {author} {\bibfnamefont {B.~R.}\ \bibnamefont {Safdi}}, \ and\ \bibinfo
  {author} {\bibfnamefont {Z.}~\bibnamefont {Sun}},\ }\href {\doibase
  10.1103/PhysRevLett.121.241102} {\bibfield  {journal} {\bibinfo  {journal}
  {Phys. Rev. Lett.}\ }\textbf {\bibinfo {volume} {121}},\ \bibinfo {pages}
  {241102} (\bibinfo {year} {2018})}\BibitemShut {NoStop}%
\bibitem [{\citenamefont {Sen}(2018)}]{srimoyee_2018}%
  \BibitemOpen
  \bibfield  {author} {\bibinfo {author} {\bibfnamefont {S.}~\bibnamefont
  {Sen}},\ }\href {\doibase 10.1103/PhysRevD.98.103012} {\bibfield  {journal}
  {\bibinfo  {journal} {Phys. Rev. D}\ }\textbf {\bibinfo {volume} {98}},\
  \bibinfo {pages} {103012} (\bibinfo {year} {2018})}\BibitemShut {NoStop}%
\bibitem [{\citenamefont {Chen}(2012)}]{9781475704617}%
  \BibitemOpen
  \bibfield  {author} {\bibinfo {author} {\bibfnamefont {F.~F.}\ \bibnamefont
  {Chen}},\ }\href
  {https://www.amazon.com/Introduction-Plasma-Physics-Francis-Chen/dp/1475704615?SubscriptionId=0JYN1NVW651KCA56C102&tag=techkie-20&linkCode=xm2&camp=2025&creative=165953&creativeASIN=1475704615}
  {\emph {\bibinfo {title} {Introduction to Plasma Physics}}}\ (\bibinfo
  {publisher} {Springer},\ \bibinfo {year} {2012})\BibitemShut {NoStop}%
\bibitem [{\citenamefont {Asseo}\ \emph {et~al.}(1980)\citenamefont {Asseo},
  \citenamefont {Pellat},\ and\ \citenamefont {Rosado}}]{asseo_1980}%
  \BibitemOpen
  \bibfield  {author} {\bibinfo {author} {\bibfnamefont {E.}~\bibnamefont
  {Asseo}}, \bibinfo {author} {\bibfnamefont {R.}~\bibnamefont {Pellat}}, \
  and\ \bibinfo {author} {\bibfnamefont {M.}~\bibnamefont {Rosado}},\ }\href
  {\doibase 10.1086/158153} {\bibfield  {journal} {\bibinfo  {journal} {The
  Astrophysical Journal}\ }\textbf {\bibinfo {volume} {239}},\ \bibinfo {pages}
  {661} (\bibinfo {year} {1980})}\BibitemShut {NoStop}%
\bibitem [{\citenamefont {Ter\ifmmode~\mbox{\c{c}}\else \c{c}\fi{}as}\ \emph
  {et~al.}(2018)\citenamefont {Ter\ifmmode~\mbox{\c{c}}\else \c{c}\fi{}as},
  \citenamefont {Rodrigues},\ and\ \citenamefont
  {Mendon\ifmmode~\mbox{\c{c}}\else \c{c}\fi{}a}}]{tercas_2018}%
  \BibitemOpen
  \bibfield  {author} {\bibinfo {author} {\bibfnamefont {H.}~\bibnamefont
  {Ter\ifmmode~\mbox{\c{c}}\else \c{c}\fi{}as}}, \bibinfo {author}
  {\bibfnamefont {J.~D.}\ \bibnamefont {Rodrigues}}, \ and\ \bibinfo {author}
  {\bibfnamefont {J.~T.}\ \bibnamefont {Mendon\ifmmode~\mbox{\c{c}}\else
  \c{c}\fi{}a}},\ }\href {\doibase 10.1103/PhysRevLett.120.181803} {\bibfield
  {journal} {\bibinfo  {journal} {Phys. Rev. Lett.}\ }\textbf {\bibinfo
  {volume} {120}},\ \bibinfo {pages} {181803} (\bibinfo {year}
  {2018})}\BibitemShut {NoStop}%
\bibitem [{\citenamefont {Das}\ \emph {et~al.}(2008)\citenamefont {Das},
  \citenamefont {Jain}, \citenamefont {Ralston},\ and\ \citenamefont
  {Saha}}]{das_2008}%
  \BibitemOpen
  \bibfield  {author} {\bibinfo {author} {\bibfnamefont {S.}~\bibnamefont
  {Das}}, \bibinfo {author} {\bibfnamefont {P.}~\bibnamefont {Jain}}, \bibinfo
  {author} {\bibfnamefont {J.~P.}\ \bibnamefont {Ralston}}, \ and\ \bibinfo
  {author} {\bibfnamefont {R.}~\bibnamefont {Saha}},\ }\href {\doibase
  10.1007/s12043-008-0060-x} {\bibfield  {journal} {\bibinfo  {journal}
  {Pramana}\ }\textbf {\bibinfo {volume} {70}},\ \bibinfo {pages} {439}
  (\bibinfo {year} {2008})}\BibitemShut {NoStop}%
\bibitem [{\citenamefont {Visinelli}(2013)}]{visinelli_2013}%
  \BibitemOpen
  \bibfield  {author} {\bibinfo {author} {\bibfnamefont {L.}~\bibnamefont
  {Visinelli}},\ }\href {\doibase 10.1142/S0217732313501629} {\bibfield
  {journal} {\bibinfo  {journal} {Modern Physics Letters A}\ }\textbf {\bibinfo
  {volume} {28}},\ \bibinfo {pages} {1350162} (\bibinfo {year}
  {2013})}\BibitemShut {NoStop}%
\bibitem [{\citenamefont {Wilczek}(1987)}]{wilczek_1987}%
  \BibitemOpen
  \bibfield  {author} {\bibinfo {author} {\bibfnamefont {F.}~\bibnamefont
  {Wilczek}},\ }\href {\doibase 10.1103/PhysRevLett.58.1799} {\bibfield
  {journal} {\bibinfo  {journal} {Phys. Rev. Lett.}\ }\textbf {\bibinfo
  {volume} {58}},\ \bibinfo {pages} {1799} (\bibinfo {year}
  {1987})}\BibitemShut {NoStop}%
\bibitem [{\citenamefont {Anderson}\ \emph {et~al.}(2001)\citenamefont
  {Anderson}, \citenamefont {Fedele},\ and\ \citenamefont
  {Lisak}}]{anderson_2001}%
  \BibitemOpen
  \bibfield  {author} {\bibinfo {author} {\bibfnamefont {D.}~\bibnamefont
  {Anderson}}, \bibinfo {author} {\bibfnamefont {R.}~\bibnamefont {Fedele}}, \
  and\ \bibinfo {author} {\bibfnamefont {M.}~\bibnamefont {Lisak}},\ }\href
  {\doibase 10.1119/1.1407252} {\bibfield  {journal} {\bibinfo  {journal}
  {American Journal of Physics}\ }\textbf {\bibinfo {volume} {69}},\ \bibinfo
  {pages} {1262} (\bibinfo {year} {2001})}\BibitemShut {NoStop}%
\bibitem [{\citenamefont {O'Neil}\ \emph {et~al.}(1971)\citenamefont {O'Neil},
  \citenamefont {Winfrey},\ and\ \citenamefont {Malmberg}}]{oneil_1971}%
  \BibitemOpen
  \bibfield  {author} {\bibinfo {author} {\bibfnamefont {T.~M.}\ \bibnamefont
  {O'Neil}}, \bibinfo {author} {\bibfnamefont {J.~H.}\ \bibnamefont {Winfrey}},
  \ and\ \bibinfo {author} {\bibfnamefont {J.~H.}\ \bibnamefont {Malmberg}},\
  }\href {\doibase 10.1063/1.1693587} {\bibfield  {journal} {\bibinfo
  {journal} {The Physics of Fluids}\ }\textbf {\bibinfo {volume} {14}},\
  \bibinfo {pages} {1204} (\bibinfo {year} {1971})},\ \Eprint
  {http://arxiv.org/abs/https://aip.scitation.org/doi/pdf/10.1063/1.1693587}
  {https://aip.scitation.org/doi/pdf/10.1063/1.1693587} \BibitemShut {NoStop}%
\bibitem [{\citenamefont {Sharma}\ and\ \citenamefont
  {Buti}(1976)}]{sharma_1976}%
  \BibitemOpen
  \bibfield  {author} {\bibinfo {author} {\bibfnamefont {A.~S.}\ \bibnamefont
  {Sharma}}\ and\ \bibinfo {author} {\bibfnamefont {B.}~\bibnamefont {Buti}},\
  }\href {\doibase 10.1007/BF02872182} {\bibfield  {journal} {\bibinfo
  {journal} {Pramana}\ }\textbf {\bibinfo {volume} {6}},\ \bibinfo {pages}
  {329} (\bibinfo {year} {1976})}\BibitemShut {NoStop}%
\bibitem [{sup()}]{supp}%
  \BibitemOpen
  \href@noop {} {\bibinfo  {journal} {Check the Supplemental Material located
  at [url] for details on the calculation of the plasmon-axion conversion
  probability, Eq. \eqref{eq_prob1}}\ }\BibitemShut {NoStop}%
\bibitem [{com()}]{comment}%
  \BibitemOpen
\bibfield  {journal} {  }\href@noop {} {\bibinfo  {journal} {Although the
  beam-plasma instability also occurs for high-density plasmas, the produced
  axions would not decay into photons resonantly, as the plasma frequency
  largely exceeds the axion masses predicted by the models.}\ }\BibitemShut
  {NoStop}%
\bibitem [{\citenamefont {Ochelkov}\ and\ \citenamefont
  {Usov}(1984)}]{ochelkov_1984}%
  \BibitemOpen
\bibfield  {journal} {  }\bibfield  {author} {\bibinfo {author} {\bibfnamefont
  {Y.~P.}\ \bibnamefont {Ochelkov}}\ and\ \bibinfo {author} {\bibfnamefont
  {V.~V.}\ \bibnamefont {Usov}},\ }\href {https://doi.org/10.1038/309332a0}
  {\bibfield  {journal} {\bibinfo  {journal} {Nature}\ }\textbf {\bibinfo
  {volume} {309}},\ \bibinfo {pages} {332 EP } (\bibinfo {year}
  {1984})}\BibitemShut {NoStop}%
\bibitem [{\citenamefont {{Ruderman}}\ and\ \citenamefont
  {{Sutherland}}(1975)}]{ruderman_1975}%
  \BibitemOpen
  \bibfield  {author} {\bibinfo {author} {\bibfnamefont {M.~A.}\ \bibnamefont
  {{Ruderman}}}\ and\ \bibinfo {author} {\bibfnamefont {P.~G.}\ \bibnamefont
  {{Sutherland}}},\ }\href {\doibase 10.1086/153393} {\bibfield  {journal}
  {\bibinfo  {journal} {\apj}\ }\textbf {\bibinfo {volume} {196}},\ \bibinfo
  {pages} {51} (\bibinfo {year} {1975})}\BibitemShut {NoStop}%
\bibitem [{\citenamefont {Goldreich}\ and\ \citenamefont
  {Julian}(1969)}]{goldreich_1969}%
  \BibitemOpen
  \bibfield  {author} {\bibinfo {author} {\bibfnamefont {P.}~\bibnamefont
  {Goldreich}}\ and\ \bibinfo {author} {\bibfnamefont {W.~H.}\ \bibnamefont
  {Julian}},\ }\href {\doibase 10.1086/150119} {\bibfield  {journal} {\bibinfo
  {journal} {The Astrophysical Journal}\ }\textbf {\bibinfo {volume} {157}},\
  \bibinfo {pages} {869} (\bibinfo {year} {1969})}\BibitemShut {NoStop}%
\bibitem [{\citenamefont {Melrose}\ and\ \citenamefont
  {Yuen}(2016)}]{melrose_2016}%
  \BibitemOpen
  \bibfield  {author} {\bibinfo {author} {\bibfnamefont {D.~B.}\ \bibnamefont
  {Melrose}}\ and\ \bibinfo {author} {\bibfnamefont {R.}~\bibnamefont {Yuen}},\
  }\href {\doibase 10.1017/s0022377816000398} {\bibfield  {journal} {\bibinfo
  {journal} {Journal of Plasma Physics}\ }\textbf {\bibinfo {volume} {82}}
  (\bibinfo {year} {2016}),\ 10.1017/s0022377816000398}\BibitemShut {NoStop}%
\bibitem [{\citenamefont {Kennea}\ \emph {et~al.}(2013)\citenamefont {Kennea},
  \citenamefont {Burrows}, \citenamefont {Kouveliotou}, \citenamefont {Palmer},
  \citenamefont {Gö?ü?}, \citenamefont {Kaneko}, \citenamefont {Evans},
  \citenamefont {Degenaar}, \citenamefont {Reynolds}, \citenamefont {Miller},
  \citenamefont {Wijnands}, \citenamefont {Mori},\ and\ \citenamefont
  {Gehrels}}]{kennea_2013}%
  \BibitemOpen
  \bibfield  {author} {\bibinfo {author} {\bibfnamefont {J.~A.}\ \bibnamefont
  {Kennea}}, \bibinfo {author} {\bibfnamefont {D.~N.}\ \bibnamefont {Burrows}},
  \bibinfo {author} {\bibfnamefont {C.}~\bibnamefont {Kouveliotou}}, \bibinfo
  {author} {\bibfnamefont {D.~M.}\ \bibnamefont {Palmer}}, \bibinfo {author}
  {\bibfnamefont {E.}~\bibnamefont {Gö?ü?}}, \bibinfo {author} {\bibfnamefont
  {Y.}~\bibnamefont {Kaneko}}, \bibinfo {author} {\bibfnamefont {P.~A.}\
  \bibnamefont {Evans}}, \bibinfo {author} {\bibfnamefont {N.}~\bibnamefont
  {Degenaar}}, \bibinfo {author} {\bibfnamefont {M.~T.}\ \bibnamefont
  {Reynolds}}, \bibinfo {author} {\bibfnamefont {J.~M.}\ \bibnamefont
  {Miller}}, \bibinfo {author} {\bibfnamefont {R.}~\bibnamefont {Wijnands}},
  \bibinfo {author} {\bibfnamefont {K.}~\bibnamefont {Mori}}, \ and\ \bibinfo
  {author} {\bibfnamefont {N.}~\bibnamefont {Gehrels}},\ }\href
  {http://stacks.iop.org/2041-8205/770/i=2/a=L24} {\bibfield  {journal}
  {\bibinfo  {journal} {The Astrophysical Journal Letters}\ }\textbf {\bibinfo
  {volume} {770}},\ \bibinfo {pages} {L24} (\bibinfo {year}
  {2013})}\BibitemShut {NoStop}%
\bibitem [{\citenamefont {Eatough}\ \emph {et~al.}(2013)\citenamefont
  {Eatough}, \citenamefont {Falcke}, \citenamefont {Karuppusamy}, \citenamefont
  {Lee}, \citenamefont {Champion}, \citenamefont {Keane}, \citenamefont
  {Desvignes}, \citenamefont {Schnitzeler}, \citenamefont {Spitler},
  \citenamefont {Kramer}, \citenamefont {Klein}, \citenamefont {Bassa},
  \citenamefont {Bower}, \citenamefont {Brunthaler}, \citenamefont {Cognard},
  \citenamefont {Deller}, \citenamefont {Demorest}, \citenamefont {Freire},
  \citenamefont {Kraus}, \citenamefont {Lyne}, \citenamefont {Noutsos},
  \citenamefont {Stappers},\ and\ \citenamefont {Wex}}]{eatough_2013}%
  \BibitemOpen
  \bibfield  {author} {\bibinfo {author} {\bibfnamefont {R.~P.}\ \bibnamefont
  {Eatough}}, \bibinfo {author} {\bibfnamefont {H.}~\bibnamefont {Falcke}},
  \bibinfo {author} {\bibfnamefont {R.}~\bibnamefont {Karuppusamy}}, \bibinfo
  {author} {\bibfnamefont {K.~J.}\ \bibnamefont {Lee}}, \bibinfo {author}
  {\bibfnamefont {D.~J.}\ \bibnamefont {Champion}}, \bibinfo {author}
  {\bibfnamefont {E.~F.}\ \bibnamefont {Keane}}, \bibinfo {author}
  {\bibfnamefont {G.}~\bibnamefont {Desvignes}}, \bibinfo {author}
  {\bibfnamefont {D.~H. F.~M.}\ \bibnamefont {Schnitzeler}}, \bibinfo {author}
  {\bibfnamefont {L.~G.}\ \bibnamefont {Spitler}}, \bibinfo {author}
  {\bibfnamefont {M.}~\bibnamefont {Kramer}}, \bibinfo {author} {\bibfnamefont
  {B.}~\bibnamefont {Klein}}, \bibinfo {author} {\bibfnamefont
  {C.}~\bibnamefont {Bassa}}, \bibinfo {author} {\bibfnamefont {G.~C.}\
  \bibnamefont {Bower}}, \bibinfo {author} {\bibfnamefont {A.}~\bibnamefont
  {Brunthaler}}, \bibinfo {author} {\bibfnamefont {I.}~\bibnamefont {Cognard}},
  \bibinfo {author} {\bibfnamefont {A.~T.}\ \bibnamefont {Deller}}, \bibinfo
  {author} {\bibfnamefont {P.~B.}\ \bibnamefont {Demorest}}, \bibinfo {author}
  {\bibfnamefont {P.~C.~C.}\ \bibnamefont {Freire}}, \bibinfo {author}
  {\bibfnamefont {A.}~\bibnamefont {Kraus}}, \bibinfo {author} {\bibfnamefont
  {A.~G.}\ \bibnamefont {Lyne}}, \bibinfo {author} {\bibfnamefont
  {A.}~\bibnamefont {Noutsos}}, \bibinfo {author} {\bibfnamefont
  {B.}~\bibnamefont {Stappers}}, \ and\ \bibinfo {author} {\bibfnamefont
  {N.}~\bibnamefont {Wex}},\ }\href {\doibase 10.1038/nature12499} {\bibfield
  {journal} {\bibinfo  {journal} {Nature}\ }\textbf {\bibinfo {volume} {501}},\
  \bibinfo {pages} {391} (\bibinfo {year} {2013})}\BibitemShut {NoStop}%
\bibitem [{\citenamefont {Huang}\ \emph {et~al.}(2018)\citenamefont {Huang},
  \citenamefont {Kadota}, \citenamefont {Sekiguchi},\ and\ \citenamefont
  {Tashiro}}]{huang_2018}%
  \BibitemOpen
  \bibfield  {author} {\bibinfo {author} {\bibfnamefont {F.~P.}\ \bibnamefont
  {Huang}}, \bibinfo {author} {\bibfnamefont {K.}~\bibnamefont {Kadota}},
  \bibinfo {author} {\bibfnamefont {T.}~\bibnamefont {Sekiguchi}}, \ and\
  \bibinfo {author} {\bibfnamefont {H.}~\bibnamefont {Tashiro}},\ }\href
  {\doibase 10.1103/PhysRevD.97.123001} {\bibfield  {journal} {\bibinfo
  {journal} {Phys. Rev. D}\ }\textbf {\bibinfo {volume} {97}},\ \bibinfo
  {pages} {123001} (\bibinfo {year} {2018})}\BibitemShut {NoStop}%
\bibitem [{\citenamefont {Redondo}\ and\ \citenamefont
  {Ringwald}(2011)}]{redondo_2011}%
  \BibitemOpen
  \bibfield  {author} {\bibinfo {author} {\bibfnamefont {J.}~\bibnamefont
  {Redondo}}\ and\ \bibinfo {author} {\bibfnamefont {A.}~\bibnamefont
  {Ringwald}},\ }\href {\doibase 10.1080/00107514.2011.563516} {\bibfield
  {journal} {\bibinfo  {journal} {Contemporary Physics}\ }\textbf {\bibinfo
  {volume} {52}},\ \bibinfo {pages} {211} (\bibinfo {year} {2011})},\ \Eprint
  {http://arxiv.org/abs/https://doi.org/10.1080/00107514.2011.563516}
  {https://doi.org/10.1080/00107514.2011.563516} \BibitemShut {NoStop}%
\bibitem [{\citenamefont {Ehret}\ \emph {et~al.}(2010)\citenamefont {Ehret},
  \citenamefont {Frede}, \citenamefont {Ghazaryan}, \citenamefont
  {Hildebrandt}, \citenamefont {Knabbe}, \citenamefont {Kracht}, \citenamefont
  {Lindner}, \citenamefont {List}, \citenamefont {Meier}, \citenamefont
  {Meyer}, \citenamefont {Notz}, \citenamefont {Redondo}, \citenamefont
  {Ringwald}, \citenamefont {Wiedemann},\ and\ \citenamefont
  {Willke}}]{alps_2010}%
  \BibitemOpen
  \bibfield  {author} {\bibinfo {author} {\bibfnamefont {K.}~\bibnamefont
  {Ehret}}, \bibinfo {author} {\bibfnamefont {M.}~\bibnamefont {Frede}},
  \bibinfo {author} {\bibfnamefont {S.}~\bibnamefont {Ghazaryan}}, \bibinfo
  {author} {\bibfnamefont {M.}~\bibnamefont {Hildebrandt}}, \bibinfo {author}
  {\bibfnamefont {E.-A.}\ \bibnamefont {Knabbe}}, \bibinfo {author}
  {\bibfnamefont {D.}~\bibnamefont {Kracht}}, \bibinfo {author} {\bibfnamefont
  {A.}~\bibnamefont {Lindner}}, \bibinfo {author} {\bibfnamefont
  {J.}~\bibnamefont {List}}, \bibinfo {author} {\bibfnamefont {T.}~\bibnamefont
  {Meier}}, \bibinfo {author} {\bibfnamefont {N.}~\bibnamefont {Meyer}},
  \bibinfo {author} {\bibfnamefont {D.}~\bibnamefont {Notz}}, \bibinfo {author}
  {\bibfnamefont {J.}~\bibnamefont {Redondo}}, \bibinfo {author} {\bibfnamefont
  {A.}~\bibnamefont {Ringwald}}, \bibinfo {author} {\bibfnamefont
  {G.}~\bibnamefont {Wiedemann}}, \ and\ \bibinfo {author} {\bibfnamefont
  {B.}~\bibnamefont {Willke}},\ }\href {\doibase
  https://doi.org/10.1016/j.physletb.2010.04.066} {\bibfield  {journal}
  {\bibinfo  {journal} {Physics Letters B}\ }\textbf {\bibinfo {volume}
  {689}},\ \bibinfo {pages} {149 } (\bibinfo {year} {2010})}\BibitemShut
  {NoStop}%
\bibitem [{\citenamefont {De~Angelis}\ \emph {et~al.}(2007)\citenamefont
  {De~Angelis}, \citenamefont {Roncadelli},\ and\ \citenamefont
  {Mansutti}}]{angelis_2007}%
  \BibitemOpen
  \bibfield  {author} {\bibinfo {author} {\bibfnamefont {A.}~\bibnamefont
  {De~Angelis}}, \bibinfo {author} {\bibfnamefont {M.}~\bibnamefont
  {Roncadelli}}, \ and\ \bibinfo {author} {\bibfnamefont {O.}~\bibnamefont
  {Mansutti}},\ }\href {\doibase 10.1103/PhysRevD.76.121301} {\bibfield
  {journal} {\bibinfo  {journal} {Phys. Rev. D}\ }\textbf {\bibinfo {volume}
  {76}},\ \bibinfo {pages} {121301} (\bibinfo {year} {2007})}\BibitemShut
  {NoStop}%
\bibitem [{\citenamefont {Tavecchio}\ \emph {et~al.}(2012)\citenamefont
  {Tavecchio}, \citenamefont {Roncadelli}, \citenamefont {Galanti},\ and\
  \citenamefont {Bonnoli}}]{tavecchio_2012}%
  \BibitemOpen
  \bibfield  {author} {\bibinfo {author} {\bibfnamefont {F.}~\bibnamefont
  {Tavecchio}}, \bibinfo {author} {\bibfnamefont {M.}~\bibnamefont
  {Roncadelli}}, \bibinfo {author} {\bibfnamefont {G.}~\bibnamefont {Galanti}},
  \ and\ \bibinfo {author} {\bibfnamefont {G.}~\bibnamefont {Bonnoli}},\ }\href
  {\doibase 10.1103/PhysRevD.86.085036} {\bibfield  {journal} {\bibinfo
  {journal} {Phys. Rev. D}\ }\textbf {\bibinfo {volume} {86}},\ \bibinfo
  {pages} {085036} (\bibinfo {year} {2012})}\BibitemShut {NoStop}%
\bibitem [{\citenamefont {Galanti}\ and\ \citenamefont
  {Roncadelli}(2018)}]{galanti_2018}%
  \BibitemOpen
  \bibfield  {author} {\bibinfo {author} {\bibfnamefont {G.}~\bibnamefont
  {Galanti}}\ and\ \bibinfo {author} {\bibfnamefont {M.}~\bibnamefont
  {Roncadelli}},\ }\href {\doibase https://doi.org/10.1016/j.jheap.2018.07.002}
  {\bibfield  {journal} {\bibinfo  {journal} {Journal of High Energy
  Astrophysics}\ }\textbf {\bibinfo {volume} {20}},\ \bibinfo {pages} {1 }
  (\bibinfo {year} {2018})}\BibitemShut {NoStop}%
\end{thebibliography}%
\bibliographystyle{apsrev4-1}

\end{document}